# BOOTSTRAP BASED ASYMPTOTIC REFINEMENTS FOR HIGH-DIMENSIONAL NONLINEAR MODELS

by

Joel L. Horowitz
Ahnaf Rafi

Department of Economics
Northwestern University
Evanston, IL 60201
USA

February 2024

## Abstract

We consider penalized extremum estimation of a high-dimensional, possibly nonlinear model that is sparse in the sense that most of its parameters are zero but some are not. We use the SCAD penalty function, which provides model selection consistent and oracle efficient estimates under suitable conditions. However, asymptotic approximations based on the oracle model can be inaccurate with the sample sizes found in many applications. This paper gives conditions under which the bootstrap, based on estimates obtained through SCAD penalization with thresholding, provides asymptotic refinements of size $O(n^{-2})$ for the error in the rejection (coverage) probability of a symmetric hypothesis test (confidence interval) and $O(n^{-1})$ for the error in the rejection (coverage) probability of a one-sided or equal tailed test (confidence interval). The results of Monte Carlo experiments show that the bootstrap can provide large reductions in errors in rejection and coverage probabilities. The bootstrap is consistent, though it does not necessarily provide asymptotic refinements, even if some parameters are close but not equal to zero. Random-coefficients logit and probit models and nonlinear moment models are examples of models to which the procedure applies.

## BOOTSTRAP BASED ASYMPTOTIC REFINEMENTS FOR HIGH-DIMENSIONAL NONLINEAR MODELS

## 1. INTRODUCTION

This paper is about using the bootstrap to obtain asymptotic refinements for inference about the sparse but possibly high-dimensional parameter $\theta_0$ that is estimated by a thresholded version of the penalized extremum estimator

(1.1) $\tilde{\theta}_n = \arg\min_{\theta \in \Theta_n}[Q_n(\chi_n, \theta) + p_{\lambda_n}(\theta)]$.

In this equation, $\Theta_n$ is a parameter set, $\chi_n$ is a random sample of size $n$ from the distribution of a random vector, $Q_n$ is a known function such as a log-likelihood function, $p_{\lambda_n}$ is a penalty function, and $\lambda_n$ is a penalization parameter. In contrast to most of the large literature on high-dimensional estimation, we do not assume that $\theta$ is the vector of parameters of a linear or generalized linear model, the vector of coefficients of a linear combination of covariates (linear index), or a linear approximation to a nonlinear function. Instead, $Q_n$ (or $-Q_n$) is the objective function of a general extremum estimator, such as a maximum likelihood estimator; linear or nonlinear regression estimator; instrumental variables estimator of a linear or nonlinear model; or generalized method of moments (GMM) estimator. The random coefficients logit and probit models are examples of widely used models that are neither generalized linear models nor linear index models and are not easily approximated by a linear combination of functions of their covariates. A non-separable, nonlinear demand model with a possibly endogenous price variable is another example. Maximum likelihood estimation of random coefficients logit probit models and GMM estimation of a demand model are among the estimators that are accommodated by the methods presented in this paper.

If $\theta_0$ has a fixed dimension and is point identified, then conditions under which $n^{1/2}(\tilde{\theta}_n - \theta_0)$ is asymptotically normally distributed without penalization are well known. See, for example, Amemiya (1985), among many other references. However, the asymptotic normal approximation can be inaccurate with the sample sizes found in many applications. Under conditions that are satisfied in many applications, the bootstrap provides asymptotic refinements for confidence intervals and hypothesis tests. See, for example, Hall (1992) and Horowitz (2001), among other references. The resulting reductions in the differences between true and nominal coverage and rejection probabilities (errors in coverage and rejection probabilities or ECPs and ERPs) can be large. This paper gives conditions under which the same asymptotic refinements can be obtained in penalized estimation of nonlinear models. We assume that $\theta_0$ is sparse, meaning that most of its components are zero but some are non-zero. We carry out inference about a non-zero component or linear combination of non-zero components. We give conditions under which bootstrap asymptotic refinements have the same order of magnitude that they would have if they were obtained from estimation of the oracle model (the model in which it is known a priori which components are non-zero and which are zero). For example, the error in the coverage (rejection) probability of a symmetrical confidence interval (hypothesis test) is $O(n^{-2})$. We also give conditions under which the bootstrap is consistent, though it does not necessarily provide asymptotic refinements, if some non-zero components of $\theta_0$ are close but not equal to zero. Thus, there is not a risk of inconsistency due to violation of the exact sparsity assumption.

Chatterjee and Lahiri (2010, 2011) give conditions under which the bootstrap based on LASSO (Tibshirani 1996) estimators of high-dimensional linear models is consistent. Chatterjee and Lahiri (2013) give conditions under which the bootstrap provides asymptotic refinements for confidence intervals and hypothesis tests based on adaptive LASSO (ALASSO, Zou 2006) estimators of high-dimensional linear models. Das, Chatterjee, and Lahiri (2022) give conditions under which the bootstrap provides asymptotic refinements for symmetrical tests and confidence intervals in penalized high dimensional linear models with a variety of penalty functions. The asymptotic refinements provided in the present paper are of the same or higher order as those of Chatterjee and Lahiri (2013) and Das, Chatterjee, and Lahiri (2022) but are for models that may be nonlinear.

We use the SCAD penalty function (Antoniadis and Fan 2001, Fan and Li 2001), which avoids penalization bias of estimates of the non-zero components of $\theta_0$. Fan and Li (2001) and Fan and Peng (2004) give conditions under which a penalized maximum likelihood estimator with the SCAD penalty function is oracle efficient, meaning that the centered and scaled estimates of non-zero components of $\theta_0$ have the same asymptotic distribution that they would have if it were known a priori which components are zero. We consider the more general estimator (1.1).

To the best of our knowledge, this paper is the first to obtain its order of asymptotic refinements for high dimensional models that may be nonlinear. Not surprisingly, achieving these refinements requires assumptions that are stronger than those in much of the recent literature. We assume that the number of parameters is less than $n$, though it may be an increasing function of $n$, and that the number of non-zero components of $\theta_0$ is fixed as $n$ increases. These assumptions are motivated by applications in economics and other social sciences, where a model may have many parameters but few have substantial effects on the dependent variable or differences from zero that can be distinguished from random sampling error. In a random coefficients logit or probit model, for example, the parameters include the means and variances of the coefficients, but some coefficients may be non-stochastic, in which case the corresponding variances are zero. As in Chatterjee and Lahiri (2013) and Das, Chatterjee, and Lahiri (2022), we require the non-zero parameters to be sufficiently far from zero. This ensures that "large" parameters can be distinguished from "small" ones with sufficiently high probability. It is not possible to obtain asymptotic refinements of the order obtained here without making such a distinction, though the bootstrap is consistent even if some non-zero parameters are "small." Appendix A presents precise statements of our assumptions and their justifications.

The results in this paper are most closely related to those of Chatterjee and Lahiri (2013) and Das, Chatterjee and Lahiri (2022), who show that suitable versions of the residual and permutation bootstraps provide asymptotic refinements for test statistics based on a large class of penalized estimators of the coefficients of a linear mean-regression model. See, also, Das, Gregory, and Lahiri (2019). These papers obtain their results by carrying out higher-order expansions of the distributions of the relevant statistics. This paper uses a different approach. We give conditions under which a combination of penalized estimation with the SCAD penalty function and hard thresholding causes the difference between the penalized, thresholded parameter estimate and the infeasible oracle estimate to converge to zero very rapidly. Consequently, the estimate obtained from penalization and thresholding can be treated as if it were the oracle estimate. It suffices to consider only the properties of the bootstrap applied to the oracle model, which are well known. It is not necessary to carry out higher-order expansions of the distribution of the penalized estimator. Penalized estimation with the SCAD penalty function is model selection consistent without thresholding (Fan and Li 2001), but the resulting parameter estimate does not converge to the oracle estimate sufficiently rapidly to provide the asymptotic refinements obtained here.

The literature on high-dimensional estimation is very large. Here, we mention only few references that are most relevant to the present paper. Tibshirani (1996) introduced the LASSO.

Knight and Fu (2000); Candès and Tao (2007); Huang, Horowitz, and Ma (2008); Zhang and Huang (2008); Belloni and Chernozhukov (2011); and Bühlmann and van de Geer (2011) describe properties of the LASSO and related penalized estimators of linear mean- and quantile-regression models. Zou (2006) introduces the ALASSO and describes its properties. Fan and Li (2001) and Fan and Peng (2004) describe properties of SCAD-penalized least squares and maximum likelihood estimators. Belloni, Chernozhukov, and Hansen (2014); van de Geer, Bühlmann, Ritov, and Dezeure (2014); Zhang and Zhang (2014); Chernozhukov, et al. (2016); Lu, Liu, Lin, and Zhang (2017); Wang, He, and Zhu (2020); and Yu, Lin, Lu and Yu (2020) describe methods for first order asymptotic inference in high-dimensional models. Bühlmann (2015) and Dezeure, Bühlmann, and Meinshausen (2015) provide reviews. Bach (2009); Chatterjee and Lahiri (2010, 2011, 2013); Javanmard and Montanari (2018); Minnier, Tian, and Cai (2011); Camponovo (2015, 2020); Dezeure, Bühlmann, and Zhang (2017); Zhang and Cheng (2017); Wang, Van Keilegom, and Maidman (2018); Das and Lahiri (2019); Liu, Xu, and Li (2020); and Das, Chatterjee, and Lahiri (2020) describe bootstrap methods for high-dimensional estimators. Chatterjee and Lahiri (2011) apply hard thresholding to the LASSO estimator of a linear model. They show that the residual bootstrap consistently estimates the distribution of the $t$ statistic this model but do not obtain asymptotic refinements or treat nonlinear models. Meinshausen and Yu (2009) and Bühlmann and van de Geer (2011) also discuss thresholding the LASSO.

The remainder of this paper is organized as follows. Section 2 describes our method and its properties. This section also treats the case in which some parameters are close but not equal to zero. Section 3 presents the results of a Monte Carlo investigation of the numerical behavior of the method, and Section 4 presents an empirical example of the application of the method. Section 5 presents conclusions. Regularity conditions are in Appendix A. The proofs of theorems and certain auxiliary results are in Appendix B.

## 2. THE METHOD

Section 2.1 defines notation that is used in the remainder of this paper. Section 2.2 presents the bootstrap method and its properties. Section 2.3 treats the case in which some parameters are close but not equal to zero.

### 2.1 *Notation*

Let $\chi_n = \{X_i : i = 1,...,n\}$ be an independent random sample of the random vector $X$. To accommodate the possibility that the dimension of the target parameter may increase as $n$ increases, we use the notation $\theta_{0n}$ for the target parameter and $\Theta_n$ for the parameter set. Define $p_n = \dim(\theta_{0n})$. Let $A_0$ and $\bar{A}_0$, respectively, denote the sets of indices of the non-zero and zero components of $\theta_{0n}$. Let $\theta_{0n}$ have $p_0$ non-zero components. This number is fixed as $n$ increases. We assume without further loss of generality that the first $p_0$ components of $\theta_{0n}$ are the non-zero ones. Thus, $A_0 = \{1,2,...,p_0\}$, and $\bar{A}_0 = \{p_0+1,...,p_n\}$. Let $\theta_{0nA_0}$ be the $p_0 \times 1$ vector of non-zero components of $\theta_{0n}$. Then $\theta_{0n} = (\theta'_{0nA_0}, 0'_{p_n-p_0})'$, where $0_{p_n-p_0}$ denotes a $(p_n - p_0)\times 1$ vector of zeros. Denote a generic element of $\Theta_n$ by $\theta_n = (\theta_{nA_0}, \theta_{n\bar{A}_0})$. Write $Q_n(\chi_n, \theta_n)$ as $Q_n(\chi_n, \theta_{nA_0}, \theta_{n\bar{A}_0})$ when it is necessary to distinguish between components whose indices are in $A_0$ and components whose indices are in $\bar{A}_0$. Let $\theta_{nj}$ ($j = 1,..., p_n$) denote the *j*'th component of any vector $\theta_n \in \Theta_n$. The penalty parameter depends on $n$ and, therefore, is denoted by $\lambda_n$. The SCAD penalty function is

$$p_{\lambda_n}(\theta_n) = \lambda_n \sum_{j=1}^{p_n} \tilde{p}_{\lambda_n}(|\theta_{nj}|),$$

where the function $\tilde{p}_{\lambda_n}$ is defined by its derivative

$$\tilde{p}'_{\lambda_n}(v) = I(v \le \lambda_n) + \frac{(a\lambda_n - v)_+}{(a-1)\lambda_n} I(v > \lambda_n); \quad a > 2, v > 0.$$

Let $\tilde{\theta}_n$ denote the penalized extremum estimator defined in (1.1) with the SCAD penalty function. Define the thresholded estimator $\hat{\theta}_n$ as the $p_n \times 1$ vector whose $j$'th component is

(2.1) $\quad \hat{\theta}_{nj} = \tilde{\theta}_{nj} I(|\tilde{\theta}_{nj}| \ge \tau_n)$,

where $\tau_n \ll \lambda_n$ is a thresholding parameter. Define the sets

$$\hat{A}_0 = \{ j = 1,..., p_n : |\hat{\theta}_{nj}| > 0 \},$$

$$\bar{\hat{A}}_0 = \{ j = 1,..., p_n : |\hat{\theta}_{nj}| = 0 \}.$$

Let $\hat{\theta}_{n\hat{A}_0}$ and $\hat{\theta}_{n\bar{\hat{A}}_0}$, respectively, denote the vectors of non-zero and zero components of $\hat{\theta}_n$. Define

$$\Theta_n^O = \{ \theta_n \in \Theta_n : \theta_{n\bar{A}_0} = 0_{p_n - p_0} \},$$

and

$$\Theta_n^{PO} = \{ \theta_n \in \Theta_n : \hat{\theta}_{n\bar{\hat{A}}_0} = 0_{|\bar{\hat{A}}_0|} \},$$

where $|\bar{\hat{A}}_0|$ is the number of elements in $\bar{\hat{A}}_0$ and $0_{|\bar{\hat{A}}_0|}$ is a $|\bar{\hat{A}}_0| \times 1$ vector of zeros. The infeasible oracle estimator of $\theta_{0n}$ is $\hat{\theta}_n^O = (\hat{\theta}_{nA_0}^O, 0_{p_n - p_0})$, where

(2.2) $\quad \hat{\theta}_{nA_0}^O = \arg\min_{\theta_n \in \Theta_n^O} Q_n(\chi_n, \theta_n)$.

This is the estimator obtained by setting $\theta_{0\bar{A}_0} = 0_{p_n - p_0}$ and choosing $\theta_{nA_0}$ to maximize the unpenalized objective function $Q_n$. Define the pseudo-oracle estimator $\hat{\theta}^{PO}_{n\hat{A}_0}$ by

(2.3) $$\hat{\theta}^{PO}_{n\hat{A}_0} = \arg \min_{\theta_n \in \in \Theta_n^{PO}} Q_n(\chi_n, \theta_n).$$

This is the estimator obtained by setting $\theta_{0\bar{\hat{A}}_0} = 0_{|\bar{\hat{A}}_0|}$ and choosing $\theta_{n\hat{A}_0}$ to maximize the unpenalized objective function.

Finally, let $T(\hat{\theta}^{PO}_{n\hat{A}_0})$ be a statistic based on $\hat{\theta}^{PO}_{n\hat{A}_0}$ for testing a hypothesis about a smooth scalar function of $\theta_{0nA_0}$. For example, $T$ might be a symmetrical $t$ statistic for testing a hypothesis about the $j$ 'th component of $\theta_{0nA_0}$. Denote the hypothesized value of this component by $\theta_{0nA_0,j}$. Then

(2.4) $$T(\hat{\theta}^{PO}_{n\hat{A}_0}) = \frac{\left|\hat{\theta}^{PO}_{n\hat{A}_0,j} - \theta_{0nA_0,j}\right|}{s^{PO}},$$

where $s^{PO}$ is a standard error, and $\theta_{0nA_0,j}$ is the hypothesized value. Let $T(\hat{\theta}^{O}_{nA_0})$ be the same statistic based on the oracle estimate $\hat{\theta}^{O}_{nA_0}$. Then, the foregoing statistic is

$$T(\hat{\theta}^{O}_{nA_0}) = \frac{\left|\hat{\theta}^{O}_{nA_0,j} - \theta_{0nA_0,j}\right|}{s^{O}},$$

where $s^{O}$ is a standard error. Let $\hat{c}^{PO}_{\alpha}$ be the $\alpha$-level critical value of $T(\hat{\theta}^{PO}_{n\hat{A}_0})$ that is obtained by the bootstrap procedure described in Section 2.2. Let $\hat{c}^{O}_{\alpha}$ be the bootstrap critical value of $T(\hat{\theta}^{O}_{nA_0})$.

2.2 *Description and Properties of the Method*

The method proposed in this paper consists of the following steps.

Step 1: Obtain the penalized estimator $\tilde{\theta}_n$ from (1.1) with the SCAD penalty function.

Step 2: Obtain the thresholded estimator $\hat{\theta}_n$ from (2.1).

Step 3: Obtain the pseudo-oracle estimator $\hat{\theta}^{PO}_{n\hat{A}_0}$ from (2.3).

Step 4: Obtain bootstrap samples by sampling the data randomly with replacement (not the residual bootstrap). Obtain the critical value $\hat{c}^{PO}_{\alpha}$ by using the conventional bootstrap methods with $\hat{\theta}^{PO}_{n\hat{A}_0}$ treated as if it were true oracle estimator $\hat{\theta}_{0nA_0}$.

If $A_0$ were known and the values of the parameters were fixed, $\hat{c}^{O}_{\alpha}$ could be obtained by applying conventional bootstrap methods such as those described by Hall (1992) and Horowitz (2001) to the oracle estimator $\hat{\theta}_{0nA_0}$. The null hypothesis being tested would be rejected at the nominal $\alpha$ level if $T(\hat{\theta}^{O}_{nA_0}) > \hat{c}^{O}_{\alpha}$. The difference between the nominal and true probabilities of rejecting a correct null hypothesis (the ERP) would be

$$ERP = \left| P\left[ T(\hat{\theta}^{O}_{nA_0}) > \hat{c}^{O}_{\alpha} \right] - \alpha \right|.$$

$ERP$ is $O(n^{-2})$ for the Studentized symmetrical statistic $T(\hat{\theta}^{O}_{nA_0})$. It is typically $O(n^{-c})$, where $c = 1/2$, $1$, $3/2$, or $2$, depending on the hypothesis and the test statistic, for non-Studentized statistics or statistics for one-sided or equal-tailed hypothesis tests. We assume that the conventional bootstrap provides the same order of refinements if some non-zero parameters approach zero at the rate specified in assumption 2(ii) of the appendix. If $\theta_{0nA_0}$ is fixed, this assumption can be replaced by the conditions of Hall (1992). Appendix A provides further discussion. This paper concentrates on the Studentized, symmetrical statistic in (2.4), but the results presented here apply after obvious modifications to the other statistics.

The method of this paper replaces the unknown $A_0$ with $\hat{A}_0$ and applies conventional bootstrap methods to $\hat{\theta}^{PO}_{n\hat{A}_0,j}$ as if $\hat{A}_0$ were non-stochastic. This works because with thresholding, $P(\hat{A}_0 \neq A_0)$ approaches zero very rapidly. The precise result is given by the following theorem.

Theorem 2.1: Let assumptions 1-5 in Appendix A hold. Then

$$(2.5)\quad P(\hat{A}_0 \neq A_0) = o(n^{-2})$$

as $n \to \infty$. In addition

$$(2.6)\quad \left| P\left[ T_n(\hat{\theta}^{PO}_{n\hat{A}_0}) > \hat{c}^{PO}_{\alpha} \right] - P\left[ T_n(\hat{\theta}^{O}_{nA_0}) > \hat{c}^{O}_{\alpha} \right] \right| = o(n^{-2}) . \blacksquare$$

It follows from Theorem 2.1 that

$$(2.7)\quad ERP = \left| P\left[ T_n(\hat{\theta}^{PO}_{n\hat{A}_0}) > \hat{c}^{PO}_{\alpha} \right] - \alpha \right| = O(n^{-2})$$

Therefore, a test based on the feasible statistic $T_n(\hat{\theta}^{PO}_{n\hat{A}_0})$ and feasible bootstrap critical value $\hat{c}^{PO}_{\alpha}$ is equivalent up to $O(n^{-2})$ to a test based on the infeasible statistic $T_n(\hat{\theta}^{O}_{nA_0})$ and infeasible bootstrap critical value $\hat{c}^{O}_{\alpha}$. Moreover, $\hat{\theta}^{PO}_{n\hat{A}_0}$ is an oracle efficient estimator of $\theta_{0n}$. A confidence interval for a smooth scalar function of $\theta_{0nA_0}$ is the set of values of the function that are not rejected by the hypothesis test. Therefore, (2.7) also applies to the error in the coverage probability (ECP) of a confidence interval.

## 2.3 *Small Parameters*

In this section, we assume that some or all components of $\theta_{0n\bar{A}_0}$ are non-zero but small in the sense that $\left\|\theta_{0n\bar{A}_0}\right\|_1 = o(\tau_n)$. We use the following additional notation. Let $\theta_{0nA_0}$ denote $p_0 \times 1$ vector of components of $\theta_{0n}$ that are "large" in the sense that $|\theta_{0nA_0,j}| \gg \lambda_n$ for all $j = 1,..., p_0$. Define $\hat{\theta}_{nA_0}^{PT}$ to be the parameter estimate obtained from the unpenalized pseudo-true model

$$\hat{\theta}_{nA_0}^{PT} = \arg\min_{\theta_{nA_0}} Q_n(\chi_n, \theta_{nA_0}, 0_{\bar{A}_0}).$$

This is the pseudo-true estimate of the large parameters that would be obtained if $A_0$ were known and all the parameters in $\bar{A}_0$ were set equal to zero.

The following theorem gives conditions under which the bootstrap consistently estimates the asymptotic distribution of

$$T(\hat{\theta}_{nA_0}^{PT}) = \frac{\hat{\theta}_{nA_0,j}^{PT} - \theta_{0nA_0,j}}{s^{PT}},$$

where $s^{PT}$ is a standard error. $T(\hat{\theta}_{nA_0}^{PT})$ is a $t$ statistic for testing a hypothesis about $\theta_{0nA_0,j}$ (the true parameter value, not a pseudo-true value) when $A_0$ is known. Under the assumptions of the theorem, the bootstrap estimates the distribution of $T(\hat{\theta}_{nA_0}^{PT})$ consistently when $A_0$ is unknown. Equivalently, the bootstrap provides a confidence interval for $\theta_{0nA_0,j}$ with asymptotically correct coverage probability.

<u>Theorem 2.2</u>: Let assumptions 1-4, 1S, and 2S of Appendix A hold. Let $\hat{\theta}_{n\hat{A}_0}^{PT}$ be the estimate of $\theta_{0nA_0}$ obtained from the unpenalized pseudo-true model with $\hat{A}_0$ in place of $A_0$:

$$\hat{\theta}_{n\hat{A}_0}^{PT} = \arg\min_{\theta_{n\hat{A}_0}} Q_n(\chi_n, \theta_{n\hat{A}_0}, 0_{\hat{\bar{A}}_0}),$$

where $\hat{A}_0$ is as defined following (2.1). Denote the bootstrap sample by $\chi_n^*$ and the bootstrap estimate based on the unpenalized pseudo-true model by

$$\theta_{n\hat{A}_0}^{*PT} = \arg\min_{\theta_{n\hat{A}_0}} Q_n(\chi_n^*, \theta_{n\hat{A}_0}, 0_{\hat{\bar{A}}_0}).$$

Define the statistic $T$ based on $\theta_{n\hat{A}_0}^{*PT}$ by

$$T(\theta_{n\hat{A}_0}^{*PT}) = \frac{\theta_{n\hat{A}_0,j}^{*PT} - \hat{\theta}_{n\hat{A}_0,j}^{PT}}{s^{*PT}},$$

where $s^{*PT}$ is the bootstrap standard error obtained when $\hat{A}_0$ is replaced with $A_0$. Then

$$\sup_z \left| P^*[T(\theta_{n\hat{A}_0}^{*PT}) \le z] - P[T(\hat{\theta}_{nA_0}^{PT}) \le z] \right| \to^p 0. \quad \blacksquare$$

## 3. MONTE CARLO EXPERIMENTS

This section reports the results of a Monte Carlo investigation of the finite-sample performance of the penalization and thresholding method. Section 3.1 describes the computational algorithm. Section 3.2 describes the investigation.

### 3.1 *Computational Algorithm*

The algorithm estimates $\theta_{0n}$ iteratively. Let $\tilde{\theta}_n^0$ denote the starting value, $t = 1, 2, ...$ index iterations, and $\tilde{\theta}_n^t$ denote the estimate of $\theta_{0n}$ at iteration $t = 1, 2, ...$. We use Zou's and Li's (2008) local linear approximation of the SCAD penalty function. We obtain $\tilde{\theta}_n^{t+1}$ from $\tilde{\theta}_n^t$ by using the coordinate descent method of Friedman, *et. al.* (2007) and Friedman, Hastie, and Tibshirani 2010) to solve

$$\tilde{\theta}_n^{t+1} = \arg\min_{\theta_n \in \Theta_n} Q_n(\theta_n) + \sum_{j=1}^{p_n} p'_\lambda(|\tilde{\theta}_n^t|)|\theta_{nj}|.$$

To speed the computation, we iterate only over non-zero components at a given $t$ and implement updates to the set of non-zero components as in Zhao, Liu, and Zhang (2018) and Ge *et al.* (2019).

3.2 *Monte Carlo Experiments*

We estimate the parameter $\theta_{0n}$ of the binary logit model

$$P(Y=1|X) = \frac{\exp(\theta'_{0n}X)}{1+\exp(\theta'_{0n}X)},$$

where $Y \in \{0,1\}$; $X \sim N(0_{p_n}, \Sigma_n)$; $\Sigma_{n,j\ell} = 0.3^{|j-\ell|}$ $(j,\ell = 1,...,p_n)$; $p_n = n/10$, $p_n = n/2$, or $p_n = 3n/4$; $p_0 = 15$; $\theta_{0n} = p_n \times 1$; and

$$\theta'_{0n} = (4,-1.5,-3,1.9,2.6,4,-1.5,-3,1.9,2.6,4,-1.5,-3,1.9,2.6,0...,0).$$

$-Q_n(\chi_n,\theta_n)$ is the log-likelihood function for estimating $\theta_n$. The penalty parameter was selected to minimize the BIC criterion

$$\min_\lambda Q_n(\chi_n,\theta) + C_n |\mathcal{S}_\lambda| \log(n),$$

where $\mathcal{S}_\lambda$ is the set of indices of non-zero components of $\theta$ chosen by penalized MLE with penalty parameter $\lambda$ but without thresholding, $|\mathcal{S}_\lambda|$ is the cardinality of $\mathcal{S}_\lambda$, and $C_n = 1$ or $\log\log p_n$ The concavity parameter in the SCAD penalty function was $a = 3.7$ (Fan and Li 2001), and the thresholding parameter was $\tau_n = n^{-1/8} a\lambda_n$. There were 500 Monte Carlo replications and 2000 bootstrap replications per experiment.

Tables 1-6 show empirical coverage probabilities of nominal one-sided and symmetrical 0.90 intervals for $\theta_{0n,1}$ and $\theta_{0n,2}$ obtained the following six ways:

1. Unpenalized maximum likelihood estimation (MLE) of the full model ( $p_n$ parameters estimated without penalization) with first-order asymptotic critical values.
2. Unpenalized MLE of the full model with critical values obtained by the bootstrap.
3. Unpenalized MLE of the infeasible oracle model ( $p_0$ parameters estimated without penalization) with first-order asymptotic critical values.
4. Unpenalized MLE of the infeasible oracle model with critical values obtained by the bootstrap.
5. Penalized and thresholded MLE with first-order asymptotic critical values.
6. Penalized and thresholded MLE with critical values obtained by the bootstrap.

The tables show that the empirical coverage probabilities obtained with the full model are far from the nominal coverage probabilities with either first-order asymptotic or bootstrap-based critical values. The empirical coverage probabilities are especially far from the nominal probabilities when $p_n = n/2$ and $p_n = 3n/4$. This is because the estimated Hessian and outer product matrices with these values of $p_n$ are nearly singular. Consequently, random sampling errors in the inverse of the estimated Hessian (or outer product), which is used for Studentization, are very large. The empirical coverage probabilities obtained with the oracle model and bootstrap-based critical values are close to the nominal probabilities, but the oracle model is unknown and infeasible in applications. The empirical coverage probabilities obtained with the penalized, thresholded estimator and bootstrap-based critical values are close to the nominal probabilities and are not sensitive to the choice of $C_n$ when $n \geq 1000$.

## 4. EMPIRICAL EXAMPLE

Gentzkow, Shapiro, and Taddy (2019) investigated the relation between party affiliation and two-word phrases (bigrams) spoken by members of Congress. We use a subset of their data to estimate a binary logit model of the probability of a member's party affiliation (Democrat or Republican) conditional on phrases that the member has used. Our data consist of observations on 4319 members of Congress from 2001-2016. The covariates are the number of times a member used each of 2441 phrases as well as 60 variables describing characteristics of the member (e.g. state represented, gender). Thus, the logit model has 2501 covariates in total. Most of the phrases are used roughly equally often by Democrats and Republicans. These phrases are unlikely to be useful for predicting party affiliation, thereby justifying an assumption of sparsity or approximate sparsity.

The model is as in (3.1), where $Y = 1$ if a member is a Republican, $Y = 0$ if the member is a Democrat, and $X$ is the $2501 \times 1$ vector of covariates. The SCAD penalty and thresholding parameters were selected as described in Section 3.1. Penalized estimation with $C_n = 1$ in (3.2) resulted in selection of 106 phrases out of the initial 2441. Penalized estimation with $C_n = \log(\log p_n)$ resulted in selection of 59 phrases, 56 of which are among the 106 selected with $C_n = 1$.

Table 7 shows point estimates of the coefficients of several example phrases and nominal 90% first-order asymptotic and bootstrap-based confidence intervals for the coefficients. The point estimates and confidence intervals are not highly sensitive to the choice of $C_n$.

## 5. CONCLUSIONS

Empirical research in economics, among other fields, often involves estimation of the parameters of a nonlinear model in which the number of parameters may be a large fraction of the sample size but most parameters are zero or close to zero. In such settings, the accuracy of inference about large parameters can be improved greatly through the use of penalized estimation methods that reduce the number of parameters that must be estimated. However, inference is usually based on asymptotic approximations that can be highly inaccurate in finite samples. Under suitable conditions, the bootstrap provides asymptotic refinements that increase the accuracy of inference, but the usual conditions, such as those of Hall (1992), are not satisfied in penalized estimation. This paper has described a method for obtaining bootstrap asymptotic refinements in penalized estimation of nonlinear models. The refinements are of the same order as those that would be achieved with the oracle model if it were known. The results of Monte Carlo experiments show that with samples of the sizes encountered in much applied research, the method can achieve large reductions in the errors of the coverage probabilities of confidence intervals. The bootstrap is consistent even if the sparsity assumption needed to obtain the asymptotic refinements reported here is not satisfied. An empirical example has illustrated the method's practical usefulness.

Table 1: Logit Coverage Probabilities for $\theta_{0n,1}$: Nominal level 0.90 $p_n = n/10$

| $n$ | Interval | Full Model Asymp. | Full Model Boot. | Oracle Model Asymp. | Oracle Model Boot. | Pseudo Oracle Model Asymp., $C_n = 1$ | Pseudo Oracle Model Boot., $C_n = 1$ | Pseudo Oracle Model Asymp. $C_n = \log\log p_n$ | Pseudo Oracle Model Boot., $C_n = \log\log p_n$ |
|---|---|---|---|---|---|---|---|---|---|
| 500 | Lower 1-Sided | 0.982 | 0.850 | 0.688 | 0.892 | 0.678 | 0.884 | 0.708 | 0.894 |
| 1000 | | 0.990 | 0.880 | 0.782 | 0.938 | 0.764 | 0.93 | 0.782 | 0.938 |
| 2000 | | 0.976 | 0.842 | 0.820 | 0.906 | 0.812 | 0.906 | 0.820 | 0.906 |
| 4000 | | 0.984 | 0.852 | 0.842 | 0.898 | 0.838 | 0.894 | 0.840 | 0.898 |
| | | | | | | | | | |
| 500 | Upper 1-Sided | 1.000 | 0.830 | 0.988 | 0.902 | 0.964 | 0.882 | 0.912 | 0.818 |
| 1000 | | 1.000 | 0.886 | 0.970 | 0.902 | 0.972 | 0.902 | 0.970 | 0.902 |
| 2000 | | 1.000 | 0.824 | 0.956 | 0.886 | 0.956 | 0.888 | 0.956 | 0.886 |
| 4000 | | 1.000 | 0.834 | 0.912 | 0.872 | 0.912 | 0.872 | 0.912 | 0.872 |
| | | | | | | | | | |
| 500 | Symmetrical | 0.982 | 0.780 | 0.796 | 0.890 | 0.764 | 0.866 | 0.734 | 0.828 |
| 1000 | | 0.990 | 0.852 | 0.882 | 0.938 | 0.870 | 0.930 | 0.882 | 0.938 |
| 2000 | | 0.976 | 0.820 | 0.878 | 0.912 | 0.876 | 0.910 | 0.878 | 0.912 |
| 4000 | | 0.984 | 0.824 | 0.882 | 0.898 | 0.878 | 0.896 | 0.882 | 0.898 |

Table 2: Logit Coverage Probabilities for $\theta_{0n,1}$: Nominal level 0.90 $p_n = n/2$

| $n$ | Interval | Full Model Asymp. | Full Model Boot. | Oracle Model Asymp. | Oracle Model Boot. | Pseudo Oracle Model Asymp., $C_n = 1$ | Pseudo Oracle Model Boot., $C_n = 1$ | Pseudo Oracle Model Asymp. $C_n = \log\log p_n$ | Pseudo Oracle Model Boot., $C_n = \log\log p_n$ |
|---|---|---|---|---|---|---|---|---|---|
| 500 | Lower 1-Sided | 1.000 | 0 | 0.664 | 0.906 | 0.572 | 0.798 | 0.730 | 0.918 |
| 1000 | | 1.000 | 0 | 0.778 | 0.906 | 0.732 | 0.876 | 0.778 | 0.912 |
| 2000 | | 1.000 | 0 | 0.812 | 0.916 | 0.786 | 0.902 | 0.812 | 0.916 |
| 4000 | | 1.000 | 0 | 0.858 | 0.908 | 0.844 | 0.896 | 0.858 | 0.910 |
| | | | | | | | | | |
| 500 | Upper 1-Sided | 1.000 | 1.000 | 0.968 | 0.884 | 0.954 | 0.878 | 0.780 | 0.692 |
| 1000 | | 1.000 | 1.000 | 0.950 | 0.884 | 0.954 | 0.896 | 0.948 | 0.888 |
| 2000 | | 1.000 | 1.000 | 0.972 | 0.914 | 0.974 | 0.918 | 0.972 | 0.916 |
| 4000 | | 1.000 | 1.000 | 0.938 | 0.880 | 0.940 | 0.882 | 0.938 | 0.882 |
| | | | | | | | | | |
| 500 | Symmetrical | 1.000 | 0 | 0.806 | 0.892 | 0.688 | 0.772 | 0.668 | 0.738 |
| 1000 | | 1.000 | 0 | 0.836 | 0.902 | 0.812 | 0.868 | 0.834 | 0.898 |
| 2000 | | 1.000 | 0 | 0.896 | 0.918 | 0.878 | 0.912 | 0.896 | 0.916 |
| 4000 | | 1.000 | 0 | 0.898 | 0.912 | 0.890 | 0.902 | 0.898 | 0.920 |

Table 3: Logit Coverage Probabilities for $\theta_{0n,1}$: Nominal level 0.90 $p_n = 3n/4$

| $n$ | Interval | Full Model Asymp. | Full Model Boot. | Oracle Model Asymp. | Oracle Model Boot. | Pseudo Oracle Model Asymp., $C_n = 1$ | Pseudo Oracle Model Boot., $C_n = 1$ | Pseudo Oracle Model Asymp., $C_n = \log\log p_n$ | Pseudo Oracle Model Boot. $C_n = \log\log p_n$ |
|---|---|---|---|---|---|---|---|---|---|
| 500 | Lower 1-Sided | 1.000 | 0 | 0.660 | 0.886 | 0.558 | 0.760 | 0.740 | 0.894 |
| 1000 | | 1.000 | 0 | 0.702 | 0.904 | 0.656 | 0.850 | 0.702 | 0.904 |
| 2000 | | 1.000 | 0 | 0.818 | 0.894 | 0.786 | 0.874 | 0.816 | 0.898 |
| 4000 | | 1.000 | 0 | 0.832 | 0.904 | 0.822 | 0.892 | 0.874 | 0.910 |
| | | | | | | | | | |
| 500 | Upper 1-Sided | 1.000 | 1.000 | 0.970 | 0.894 | 0.958 | 0.902 | 0.726 | 0.618 |
| 1000 | | 1.000 | 1.000 | 0.898 | 0.898 | 0.976 | 0.904 | 0.970 | 0.904 |
| 2000 | | 1.000 | 1.000 | 0.900 | 0.900 | 0.974 | 0.916 | 0.962 | 0.902 |
| 4000 | | 1.000 | 1.000 | 0.906 | 0.906 | 0.966 | 0.916 | 0.930 | 0.905 |
| | | | | | | | | | |
| 500 | Symmetrical | 1.000 | 0 | 0.798 | 0.880 | 0.660 | 0.748 | 0.584 | 0.654 |
| 1000 | | 1.000 | 0 | 0.830 | 0.894 | 0.768 | 0.846 | 0.828 | 0.896 |
| 2000 | | 1.000 | 0 | 0.868 | 0.904 | 0.856 | 0.884 | 0.868 | 0.906 |
| 4000 | | 1.000 | 0 | 0.874 | 0.902 | 0.862 | 0.892 | 0.874 | 0.896 |

Table 4: Logit Coverage Probabilities for $\theta_{0n,2}$ : Nominal level 0.90 $p_n = n/10$

| $n$ | Interval | Full Model Asymp. | Full Model Boot. | Oracle Model Asymp. | Oracle Model Boot. | Pseudo Oracle Model Asymp., $C_n = 1$ | Pseudo Oracle Model Boot., $C_n = 1$ | Pseudo Oracle Model Asymp. $C_n = \log\log p_n$ | Pseudo Oracle Model Boot., $C_n = \log\log p_n$ |
|---|---|---|---|---|---|---|---|---|---|
| 500 | Lower 1-Sided | 0.996 | 0.810 | 0.958 | 0.91 | 0.948 | 0.908 | 0.886 | 0.864 |
| 1000 | | 0.994 | 0.824 | 0.972 | 0.916 | 0.974 | 0.916 | 0.972 | 0.916 |
| 2000 | | 0.996 | 0.818 | 0.938 | 0.902 | 0.940 | 0.902 | 0.938 | 0.902 |
| 4000 | | 0.994 | 0.778 | 0.916 | 0.894 | 0.918 | 0.894 | 0.916 | 0.894 |
| | | | | | | | | | |
| 500 | Upper 1-Sided | 0.994 | 0.842 | 0.760 | 0.924 | 0.726 | 0.890 | 0.672 | 0.822 |
| 1000 | | 0.994 | 0.806 | 0.816 | 0.938 | 0.810 | 0.932 | 0.816 | 0.938 |
| 2000 | | 0.986 | 0.814 | 0.836 | 0.908 | 0.836 | 0.908 | 0.836 | 0.908 |
| 4000 | | 0.988 | 0.832 | 0.870 | 0.914 | 0.866 | 0.912 | 0.868 | 0.912 |
| | | | | | | | | | |
| 500 | Symmetrical | 0.996 | 0.786 | 0.868 | 0.912 | 0.846 | 0.89 | 0.788 | 0.828 |
| 1000 | | 0.998 | 0.75 | 0.904 | 0.944 | 0.896 | 0.938 | 0.904 | 0.944 |
| 2000 | | 0.998 | 0.752 | 0.904 | 0.912 | 0.904 | 0.912 | 0.904 | 0.912 |
| 4000 | | 0.994 | 0.762 | 0.902 | 0.912 | 0.900 | 0.912 | 0.900 | 0.912 |

Table 5: Logit Coverage Probabilities for $\theta_{0n,2}$: Nominal level 0.90 $p_n = n/2$

| $n$ | Interval | Full Model Asymp. | Full Model Boot. | Oracle Model Asymp. | Oracle Model Boot. | Pseudo Oracle Model Asymp., $C_n = 1$ | Pseudo Oracle Model Boot., $C_n = 1$ | Pseudo Oracle Model Asymp. $C_n = \log\log p_n$ | Pseudo Oracle Model Boot., $C_n = \log\log p_n$ |
|---|---|---|---|---|---|---|---|---|---|
| 500 | Lower 1-Sided | 1.000 | 0.948 | 0.708 | 0.902 | 0.602 | 0.824 | 0.638 | 0.806 |
| 1000 | | 1.000 | 0.992 | 0.784 | 0.904 | 0.748 | 0.888 | 0.782 | 0.900 |
| 2000 | | 1.000 | 1.000 | 0.830 | 0.910 | 0.814 | 0.898 | 0.830 | 0.910 |
| 4000 | | 1.000 | 1.000 | 0.814 | 0.874 | 0.810 | 0.864 | 0.814 | 0.874 |
| | | | | | | | | | |
| 500 | Upper 1-Sided | 1.000 | 0.052 | 0.972 | 0.908 | 0.984 | 0.918 | 0.912 | 0.858 |
| 1000 | | 1.000 | 0.008 | 0.956 | 0.902 | 0.956 | 0.910 | 0.956 | 0.902 |
| 2000 | | 1.000 | 0 | 0.954 | 0.902 | 0.956 | 0.910 | 0.954 | 0.902 |
| 4000 | | 1.000 | 0 | 0.942 | 0.904 | 0.948 | 0.906 | 0.942 | 0.904 |
| | | | | | | | | | |
| 500 | Symmetrical | 1.000 | 0 | 0.808 | 0.902 | 0.712 | 0.826 | 0.662 | 0.750 |
| 1000 | | 1.000 | 0 | 0.80 | 0.908 | 0.834 | 0.888 | 0.846 | 0.904 |
| 2000 | | 1.000 | 0 | 0.892 | 0.916 | 0.882 | 0.902 | 0.892 | 0.916 |
| 4000 | | 1.000 | 0 | 0.870 | 0.880 | 0.866 | 0.872 | 0.870 | 0.880 |

Table 6: Logit Coverage Probabilities for $\theta_{0n,2}$: Nominal level 0.90 $p_n = 3n/4$

| $n$ | Interval | Full Model Asymp. | Full Model Boot. | Oracle Model Asymp. | Oracle Model Boot. | Pseudo Oracle Model Asymp., $C_n = 1$ | Pseudo Oracle Model Boot., $C_n = 1$ | Pseudo Oracle Model Asymp. $C_n = \log\log p_n$ | Pseudo Oracle Model Boot., $C_n = \log\log p_n$ |
|---|---|---|---|---|---|---|---|---|---|
| 500 | Lower 1-Sided | 1.000 | 0.948 | 0.692 | 0.922 | 0.051 | 0.788 | 0.672 | 0.844 |
| 1000 | | 1.000 | 0.992 | 0.768 | 0.924 | 0.696 | 0.882 | 0.768 | 0.924 |
| 2000 | | 1.000 | 1.000 | 0.796 | 0.892 | 0.772 | 0.896 | 0.796 | 0.892 |
| 4000 | | 1.000 | 1.000 | 0.862 | 0.914 | 0.846 | 0.900 | 0.862 | 0.914 |
| | | | | | | | | | |
| 500 | Upper 1-Sided | 1.000 | 0.052 | 0.974 | 0.898 | 0.982 | 0.922 | 0.916 | 0.824 |
| 1000 | | 1.000 | 0.008 | 0.948 | 0.890 | 0.958 | 0.904 | 0.946 | 0.890 |
| 2000 | | 1.000 | 0 | 0.950 | 0.904 | 0.954 | 0.920 | 0.950 | 0.904 |
| 4000 | | 1.000 | 0 | 0.936 | 0.898 | 0.940 | 0.904 | 0.936 | 0.898 |
| | | | | | | | | | |
| 500 | Symmetrical | 1.000 | 0 | 0.822 | 0.912 | 0.638 | 0.786 | 0.704 | 0.790 |
| 1000 | | 1.000 | 0 | 0.852 | 0.912 | 0.798 | 0.878 | 0.850 | 0.910 |
| 2000 | | 1.000 | 0 | 0.876 | 0.904 | 0.848 | 0.880 | 0.876 | 0.904 |
| 4000 | | 1.000 | 0 | 0.892 | 0.902 | 0.890 | 0.900 | 0.892 | 0.902 |

Table 7: Coefficients and Confidence Intervals in the Empirical Example

| Phrase | Interval | Estimated Coeff. $C_n = 1$ | Pseudo Oracle Model Asymp., $C_n = 1$ | Pseudo Oracle Model Boot., $C_n = 1$ | Estimated Coeff $C_n = \log\log p_n$ | Pseudo Oracle Model Asymp. $C_n = \log\log p_n$ | Pseudo Oracle Model \Boot., $C_n = \log\log p_n$ |
|---|---|---|---|---|---|---|---|
| Federal regulation | Lwr. 1-Sided | 0.53 | $(-\infty, 0.66)$ | $(-\infty, 0.69)$ | 0.57 | $(-\infty, 0.70)$ | $(-\infty, 0.70)$ |
| Medical liability | | 0.84 | $(-\infty, 1.10)$ | $(-\infty, 1.10)$ | 0.81 | $(-\infty, 1.03)$ | $(-\infty, 1.11)$ |
| Gun violence | | -0.64 | $(-\infty, -0.50)$ | $(-\infty, -0.42)$ | -0.58 | $(-\infty, -0.45)$ | $(-\infty, -0.40)$ |
| Death tax | | 0.61 | $(-\infty, 0.77)$ | $(-\infty, 0.91)$ | 0.71 | $(-\infty, 0.86)$ | $(-\infty, 0.99)$ |
| | | | | | | | |
| Federal regulation | Upr. 1-Sided | 0.53 | $(0.41, \infty)$ | $(0.39, \infty)$ | 0.57 | $(0.47, \infty)$ | $(0.45, \infty)$ |
| Medical liability | | 0.84 | $(0.57, \infty)$ | $(0.46, \infty)$ | 0.81 | $(0.59, \infty)$ | $(0.42, \infty)$ |
| Gun violence | | -0.64 | $(-0.79, \infty)$ | $(-0.72, \infty)$ | -0.58 | $(-0.72, \infty)$ | $(-0.67, \infty)$ |
| Death tax | | 0.61 | $(0.45, \infty)$ | $(0.29, \infty)$ | 0.71 | $(0.56, \infty)$ | $(0.41, \infty)$ |
| | | | | | | | |
| Federal regulation | Symm. | 0.53 | (0.37,0.69) | (0.34,0.73) | 0.57 | (0.44,0.71) | (0.41,0.73) |
| Medical liability | | 0.84 | (0.50,1.18) | (0.41,1.27) | 0.81 | (0.53,1.09) | (0.43,1.20) |
| Gun violence | | -0.64 | (-0.84,-0.45) | (-0.88,-0.41) | -0.58 | (-0.76-0.41) | (-0.78,-0.39) |
| Death tax | | 0.61 | (0.40,0.82) | (0.21,1.00) | 0.71 | (0.52,0.91) | (0.34,1.09) |

# APPENDIX A

This appendix presents regularity conditions for Theorems 2.1 and 2.2. The proofs of the theorems are in Appendix B.

A.1 *Assumptions for Theorem 2.1*

Assumption 1: (i) $\chi_n = \{X_i : i = 1, ..., n\}$ is an independent random sample from the distribution of the random vector $X$. (ii) $\Theta_n$ is compact. There is a constant $C < \infty$ such that $\|\theta_n\|_1 \le C$ for all $n$ and all $\theta_n \in \Theta_n$.

Assumption 2: (i) $p_n = O(n^b)$ for some $0 \le b < 1$. (ii) $|\theta_{0nA_0,j}| \gg \lambda_n$ for each $j = 1, ..., p_0$. (iii) $\lambda_n = \lambda_0 n^{-1/4+2\zeta}$, where $0 < \zeta < 1/8$ and $\lambda_0 > 0$ is a constant.

Define the following quantities.

a. $\delta_n = n^{-d}$ and $m_n = \exp(n\delta_n)$, where $1 - 4\zeta < d < 1 - b$.

b. $\tau_n = \tau_0 n^{-1/4+\zeta}$, where $\tau_0$ is a constant and $0 < \tau_0 < a\lambda_0$.

c. The set $\mathcal{V}_n = \left\{\theta_n : \|\theta_n - \theta_{0n}\|_1 \le v\right\}$, where $v > 0$ is a constant.

d. $S_{n\infty}(\theta_n) = Q_{n\infty}(\theta_n) + p_{\lambda_n}(\theta_n)$, where $Q_{n\infty}$ is the non-stochastic function defined in assumption 3(i) below and $p_{\lambda_n}$ is the SCAD penalty function.

e. $\eta_{nv} = \inf_{\|\theta_n - \theta_{0n}\|_1 > v} |S_{n\infty}(\theta_n) - S_{n\infty}(\theta_{0n})|$.

Assumption 3: (i) There are a non-stochastic function $Q_{n\infty}(\theta_n)$ and positive, finite constants $A$, $\varepsilon_0$, $c$, $\ell$, and $n_0$ such that

$$P\left[\sup_{\theta_n \in \Theta_n} |Q_n(\chi_n, \theta_n) - Q_{n\infty}(\theta_n)| > \varepsilon\right] \le Am_n \exp(-cn\varepsilon^2)$$

for any $\ell\tau_n^2 \le \varepsilon \le \varepsilon_0$ and $n > n_0$. (ii) For each $n$ such that $p_n > p_0$, $Q_{n\infty}(\theta_n)$ has a unique global minimum in $\Theta_n$ at a point $(\theta'_{0nA_0}, 0'_{p_n-p_0})' \in \text{int}(\Theta_n)$. (iii) In a neighborhood of $(\theta'_{0nA_0}, 0'_{p_n-p_0})'$, $Q_{n\infty}(\theta_n)$ is convex, strictly increasing in $\|\theta_n - \theta_{0n}\|_1$, and twice continuously differentiable. (iv) There is a constant $\rho > 0$ such that for any $v > 0$ and any $n$, $\inf_{\|\theta_{nA_0} - \theta_{0nA_0}\|_2 \ge v}[Q_{n\infty}(\theta_n) - Q_{n\infty}(\theta_{0n})] \ge \rho v^2$.

Define

$$H_{n,11}(\theta_n) = \frac{\partial^2 Q_{n\infty}(\theta_n)}{\partial\theta_{nA_0}\partial\theta'_{nA_0}},$$

$$H_{n,12}(\theta_n) = \frac{\partial^2 Q_{n\infty}(\theta_n)}{\partial\theta_{nA_0}\partial\theta'_{n\bar{A}_0}},$$

$$H_{n,21}(\theta_n) = H_{n,12}(\theta_n)',$$

and

$$H_{n,22}(\theta_n) = \frac{\partial^2 Q_{n\infty}(\theta_n)}{\partial\theta_{n\bar{A}_0}\partial\theta'_{n\bar{A}_0}}.$$

Note that $H_{n,12}$ is a $p_0 \times (p_n - p_0)$ matrix. Let $\mu_n(\theta_n)$ denote the smallest eigenvalue of $H_{n,11}(\theta_n)$.

Assumption 4: There is a $v > 0$ such that for all $\theta_n \in \mathcal{V}_n$: (i) $\mu_n(\theta_n) \ge \mu_0$ for all $n$ and some $\mu_0 > 0$. (ii) The components of $H_{n21}(\theta_n)H_{n11}^{-1}(\theta_n)$ are bounded for all $n$.

Assumption 5: The bootstrap provides asymptotic refinements through $O(n^{-2})$ for $P[T(\hat{\theta}_{nA_0}^O) > \hat{c}_\alpha^O]$. That is, $|P[T(\hat{\theta}_{nA_0}^O) > \hat{c}_\alpha^O] - \alpha| = o(n^{-2})$.

Assumption 1 specifies the sampling process and parameter set. The requirement that $\|\theta_n\|_1 \leq C$ is not restrictive in practice because $\theta_n$ has only finitely many non-zero components. Assumption 2(i) restricts the rate at which $p_n$ can grow as $n$ increases and rules out $p_n > n$. Chatterjee and Lahiri (2013) obtain bootstrap asymptotic refinements through $O_p(n^{-1/2})$ with $p_n > n$ for a linear model. Fan and Lv (2011) give conditions under which $P(\hat{\theta}_{n\bar{A}_0} = 0) = O(n^{-1})$ for a generalized linear model with $p_n > n$. This this paper gives conditions under which $P(\hat{A}_0 = A_0) = o(n^{-2})$ and the bootstrap achieves refinements through $O(n^{-2})$ for a large class of nonlinear models that contains but is not restricted to linear and generalized linear models. Assumption 2(ii) allows the components of $\theta_{0nA_0}$ to be small, but they must be larger than random sampling error. Keeping non-zero coefficients sufficiently far from zero is necessary to obtain model selection consistency and asymptotic refinements. See, for example, Pötscher and Leeb (2009); Bühlmann and van de Geer (2011); Chatterjee and Lahiri (2011, 2013); Fan and Lv (2011); and Das, Gregory, and Lahiri (2019). Assumptions 2(iii) specifies the rate of convergence to zero of the penalization parameter. Assumption 3(i) is a high-level restriction on the objective function $Q_n$. In typical applications, $Q_{n\infty}(\theta_n) = EQ_n(\chi_n, \theta_n)$. Proposition 1 in Section B.2 of Appendix B gives conditions under which assumption 3(i) is satisfied. Section B.2 also presents examples of models and objective functions that satisfy these conditions, including log-likelihood functions and objective functions of GMM estimation. Assumptions 3(ii)-3(iv) and 4(i)-4(ii) place restrictions on the shapes of the functions $Q_{n\infty}$ and ensure that the non-zero parameter vector $\theta_{0nA_0}$ is identified. Assumption 5 applies to the oracle model and holds under conditions given by Hall (1992) if the non-zero parameters have fixed values. Thus, assumption 5 holds under Hall's (1992) conditions if "$|\theta_{0nA_0,j}| \gg \lambda_n$" in assumption 2(ii) is replaced by "$\theta_{0nA_0,j}$ is a constant."

If $\theta_{0nA_0}$ is fixed, then under Hall's (1992) conditions, $n^{1/2}(\hat{\theta}_{0nA_0} - \theta_{0nA_0})/s$ and its bootstrap analog can be approximated with sufficient accuracy by functions of sample moments whose distributions have valid Edgeworth expansions. If $|\theta_{0nA_0,j}|$ decreases for one or more $j$'s as $n$ increases, the functions of sample moments become triangular arrays. Assumption 5 is satisfied with one or more decreasing $|\theta_{0nA_0,j}|$'s if the Edgeworth expansions remain valid. The cumulants that enter the expansions and are fixed when $\theta_{0nA_0}$ is fixed depend on $n$ when $\theta_{0nA_0}$ decreases with increasing $n$.

A.2 *Additional assumptions for Theorem 2.2*

<u>Assumption 1S</u>: $\left\|\theta_{0n\bar{A}_0}\right\|_1 = o(\tau_n)$ as $n \to \infty$.

<u>Assumption 2S</u>: (i) The bootstrap consistently estimates the asymptotic distribution of $T^{PT}(\hat{\theta}^{PT}_{nA_0}) = (\hat{\theta}^{PT}_{nA_0,j} - \breve{\theta}_{0nA_0,j})/s^{PT}$, where $\breve{\theta}_{0nA_0}$ is the population pseudo-true value of $\theta_{0nA_0}$ defined in Section B.2 of Appendix B. That is,

$$\sup_z \left| P^*[T(\theta^{*PT}_{nA_0}) \le z] - P[T^{PT}(\hat{\theta}^{PT}_{nA_0}) \le z] \right| \to^p 0.$$

(ii) $P[n^{1/2}(\hat{\theta}^{PT}_{nA_0} - \breve{\theta}_{0nA_0}) \le z]$ is a continuous function of $z$.

(iii) $s^{PT}$ converges in probability to a non-stochastic, non-zero constant.

## APPENDIX B

Section B.1 presents the proof of Theorem 2.1. Section B.2 presents the proof of Theorem 2.2. Section B.3 presents auxiliary results. Section B.4 presents additional Monte Carlo results.

### B.1 PROOF OF THEOREM 2.1

Assumptions 1-5 hold throughout this section. Define

$$S_n(\chi_n,\theta_n) = Q_n(\chi_n,\theta_n) + p_{\lambda_n}(\theta_n)$$

and

$$S_{n\infty}(\theta_n) = Q_{n\infty}(\theta_n) + p_{\lambda_n}(\theta_n),$$

where $p_{\lambda_n}$ is the SCAD penalty function.

<u>Lemma B.1</u>: Given any sequence $\{\theta_n\} \in \Theta_n$ and all sufficiently large $n$, $S_{n\infty}(\theta_n) \geq S_{n\infty}(\theta_{0n})$ with equality holding for all sufficiently large $n$ only if $\theta_n = \theta_{0n}$.

<u>Proof</u>: Define $\delta_{nA_0,j} = \theta_{nA_0,j} - \theta_{0nA_0,j}$. Then,

$$S_{n\infty}(\theta_n) - S_{n\infty}(\theta_{0n}) = Q_{n\infty}(\theta_n) - Q_{n\infty}(\theta_{0n})$$

$$+ \lambda_n \sum_{j=1}^{p_0} [\tilde{p}_{\lambda_n}(|\theta_{0nA_0,j} + \delta_{nA_0,j}|) - \tilde{p}_{\lambda_n}(|\theta_{0nA_0,j}|)] + \lambda_n \sum_{j=p_0+1}^{p_n} \tilde{p}_{\lambda_n}(|\theta_{n\bar{A}_0,j}|)$$

$$\geq Q_{n\infty}(\theta_n) - Q_{n\infty}(\theta_{0n}) - 0.5 p_0 (1+a)\lambda_n^2 + \lambda_n \sum_{j=p_0+1}^{p_n} \tilde{p}_{\lambda_n}(|\theta_{n\bar{A}_0,j}|).$$

If $|\theta_{nA_0,j} - \theta_{0nA_0,j}| \gg \lambda_n$ for some $j = 1,..., p_0$, then $[Q_{n\infty}(\theta_n) - Q_{n\infty}(\theta_{0n})| \gg \rho\lambda_n^2$ for all sufficiently large $n$ by assumption 3(iv). Therefore,

$$S_{n\infty}(\theta_n) - S_{n\infty}(\theta_{0n}) \gg \lambda_n^2 + \lambda_n \sum_{j=p_0+1}^{p_n} \tilde{p}_{\lambda_n}(|\theta_{n\bar{A}_0,j}|),$$

and $S_{n\infty}(\theta_n) - S_{n\infty}(\theta_{0n}) > 0$ for all sufficiently large $n$. If $|\theta_{nA_0,j} - \theta_{0nA_0,j}| \le c\lambda_n$ for some $c > 0$, then by assumption 2(ii)

$$\tilde{p}_{\lambda_n}(|\theta_{nA_0,j}|) - \tilde{p}_{\lambda_n}(|\theta_{0nA_0,j}|) = 0$$

for all $j = 1,..., p_0$ and sufficiently large $n$. Therefore,

$$S_{n\infty}(\theta_n) - S_{n\infty}(\theta_{0n}) = Q_{n\infty}(\theta_n) - Q_{n\infty}(\theta_{0n}) + \lambda_n \sum_{j=p_0+1}^{p_n} \tilde{p}_{\lambda_n}(|\theta_{n\bar{A}_0,j}|)$$

$$\ge \lambda_n \sum_{j=p_0+1}^{p_n} \tilde{p}_{\lambda_n}(|\theta_{n\bar{A}_0,j}|).$$

The lemma follows from the observation that $\tilde{p}(|\theta_{n\bar{A}_0,j}|) > 0$ if $\theta_{n\bar{A}_0 j} \ne \theta_{0n\bar{A}_0 j}$. Q.E.D.

Define

$$\mathcal{N}_n = \{\theta \in \Theta_n : \|\theta - \theta_{0n}\|_1 < \tau_n\}$$

and

$$\varepsilon_n = \inf_{\theta \in \Theta_n \cap \bar{\mathcal{N}}_n} S_{n\infty}(\theta) - S_{n\infty}(\theta_{0n}).$$

It follows from lemma B.3 below that $\varepsilon_n > 0$ for all sufficiently large $n$. Let $B_n$ be the event

$$\sup_{\theta \in \Theta_n} |S_n(\chi_n, \theta) - S_{n\infty}(\theta)| < \varepsilon_n / 2.$$

Lemma B.2: $\hat{A}_0 = A_0$ for all sufficiently large $n$ if $B_n$ occurs.

Proof: It follows from the definition of $B_n$ that

(B1.1) $B_n \Rightarrow S_{n\infty}(\tilde{\theta}_n) - \varepsilon_n / 2 < S_n(\chi_n, \tilde{\theta}_n)$

and

(B1.2) $B_n \Rightarrow S_n(\chi_n, \theta_{0n}) - \varepsilon_n / 2 < S_{n\infty}(\theta_{0n})$.

By the definition of $\tilde{\theta}_n$, $S_n(\chi_n, \tilde{\theta}_n) \le S_n(\chi_n, \theta_{0n})$. Therefore,

(B1.3) $B_n \Rightarrow S_{n\infty}(\tilde{\theta}_n) - \varepsilon_n / 2 < S_n(\chi_n, \theta_{0n})$.

Combining (B1.2) and (B1.3) yields

$$B_n \Rightarrow S_{n\infty}(\tilde{\theta}_n) - \varepsilon_n < S_{n\infty}(\theta_{0n})$$

and

$$B_n \Rightarrow S_{n\infty}(\tilde{\theta}_n) - S_{n\infty}(\theta_{0n}) < \varepsilon_n .$$

Therefore,

$$B_n \Rightarrow \left\| \tilde{\theta}_n - \theta_{0n} \right\|_1 < \tau_n ,$$

$$B_n \Rightarrow \left\| \tilde{\theta}_{n\bar{A}_0} \right\|_1 < \tau_n ,$$

and

$$B_n \Rightarrow | \hat{\theta}_{nj} | > 2\tau_n$$

for all $j = 1, ..., p_0$ and sufficiently large $n$. Moreover,

$$B_n \Rightarrow \; | \tilde{\theta}_{nj} | \le \tau_n \text{ and } | \hat{\theta}_{nj} | = 0 \text{ for all } j = p_0 + 1, ..., p_n .$$

It follows from the definition of $\tau_n$ that $B_n \Rightarrow \hat{A}_0 = A_0$. Q.E.D.

Lemma B.3: There is a finite constant $C_\varepsilon$ such that

$$\varepsilon_n \ge C_\varepsilon \tau_n^2$$

for all sufficiently large $n$.

Proof: Let $\mathcal{N}_\tau = \{\theta \in \Theta_n : \left\| \theta - \theta_{0n} \right\|_1 \le \tau_n\}$. By assumption 3(iii), for all sufficiently large $n$ and, therefore, sufficiently small $\tau_n$, $S_{n\infty}(\theta)$ is a strictly increasing function of $\left\| \theta - \theta_{0n} \right\|_1$ for every $\theta \in \mathcal{N}_\tau$. Moreover, for every $\theta \in \bar{\mathcal{N}}_\tau \cap \Theta_n$

$$S_n(\theta) > \min_{\vartheta \in \Theta_n : \|\vartheta - \theta_{0n}\|_1 = \tau_n} S_n(\vartheta)$$

for all sufficiently large $n$. Therefore, $\arg \min_{\vartheta \in \Theta_n \cap \bar{\mathcal{N}}_n} S_{n\infty}(\vartheta)$ is a point $\theta$ such that $\|\theta - \theta_{0n}\|_1 = \tau_n$. As in the proof of Lemma B.1,

$$S_{n\infty}(\theta_n) - S_{n\infty}(\theta_{0n}) = Q_{n\infty}(\theta_n) - Q_{n\infty}(\theta_{0n})$$

$$+ \lambda_n \sum_{j=1}^{p_0} [\tilde{p}_{\lambda_n}(|\theta_{nA_0j}|) - \tilde{p}_{\lambda_n}(|\theta_{0nA_0,j}|)] + \lambda_n \sum_{j=p_0+1}^{p_n} \tilde{p}_{\lambda_n}(|\theta_{n\bar{A}_0,j}|).$$

Moreover, $S_n(\theta_n)$ is convex if $\|\theta_n - \theta_{0n}\|_1 = \tau_n$. But $|\theta_{0nA_0,j}| \gg \lambda_n$ and $\tau_n \ll \lambda_n$. Therefore, $\tilde{p}_{\lambda_n}(|\theta_{nA_0j}|) - \tilde{p}_{\lambda_n}(|\theta_{0nA_0,j}|) = 0$ if $\|\theta_n - \theta_{0n}\|_1 = \tau_n$. It is shown in Lemma B.4 below that $\left(\|\theta_{nA_0} - \theta_{0nA_0}\|_1 = \tau_n,\ \theta_{n\bar{A}_0} = 0\right)$ satisfies the Karush-Kuhn-Tucker (KKT) conditions for minimizing $S_{n\infty}(\theta_n) - S_{n\infty}(\theta_{0n})$ over the relevant region. Therefore, $\left(\|\theta_{nA_0} - \theta_{0nA_0}\|_1 = \tau_n,\ \theta_{n\bar{A}_0} = 0\right)$ minimizes $S_{n\infty}(\theta_n) - S_{n\infty}(\theta_{0n})$ over $\theta_n \in \Theta_n \bigcap \bar{\mathcal{N}}_n$.

Now define

$$s_j = \begin{cases} 1 \text{ if } (\theta_{nj} - \theta_{0n,j}) \geq 0 \text{ and } j \in A_0 \\ -1 \text{ if } (\theta_{nj} - \theta_{0n,j}) < 0 \text{ and } j \in A_0 \\ 0 \text{ if} j \in \bar{A}_0. \end{cases}$$

and $s = (s_1, ..., s_{p_n})'$. Then by a Taylor series expansion,

$$Q_{n\infty}(\theta_{nA_0}, 0_{p_n - p_0}) - Q(\theta_{0n}) = \frac{1}{2}(\theta_{nA_0} - \theta_{0nA_0})' H_{n11}(\breve{\theta}_n)(\theta_{nA_0} - \theta_{0nA_0})$$

$$\geq \frac{\mu_0}{2} \left\| \theta_{nA_0} - \theta_{0nA_0} \right\|_2^2$$

$$\geq \frac{\mu_0}{2p_0} \left\| \theta_{nA_0} - \theta_{0nA_0} \right\|_1^2 = \frac{\mu_0 \tau_n^2}{2p_0}$$

for all sufficiently large $n$, where $\breve{\theta}_n$ is the Taylor series intermediate point. Therefore,

$$\varepsilon_n \geq \frac{\mu_0 \tau_n^2}{2p_0}.$$

Set $C_\varepsilon = \frac{\mu_0}{2p_0}$. Q.E.D.

Lemma B.4: $\left( \left\| \theta_{A_0} - \theta_{0nA_0} \right\|_1 = \tau_n,\ \theta_{n\bar{A}_0} = 0 \right)$ satisfies the Karush-Kuhn-Tucker (KKT) conditions for minimizing $S_{n\infty}(\theta) - S_{n\infty}(\theta_{0n})$ over $\theta_n \in \Theta_n \cap \bar{\mathcal{N}}_n$ for all sufficiently large $n$.

Proof:

(B1.4) $S_{n\infty}(\theta) - S_{n\infty}(\theta_{0n}) = Q_{n\infty}(\theta) - Q_{n\infty}(\theta_{0n}) + \lambda_n \sum_{j \in A_0} [p_{\lambda_n}(\theta_j) - p_{\lambda_n}(\theta_{0n,j})] + \lambda_n \sum_{j \in \bar{A}_0} p_{\lambda_n}(\theta_j)$.

Define

$$h(\theta_n) = \left\| \theta_{nA_0} - \theta_{0nA_0} \right\|_1 = \sum_{j \in A_0} [(\theta_{nj} - \theta_{0n,j}) I(\theta_{nj} - \theta_{0n,j} \geq 0) - (\theta_{nj} - \theta_{0n,j}) I(\theta_{nj} - \theta_{0n,j} < 0)].$$

Then

$$\frac{\partial h(\theta_n)}{\partial \theta_{nj}} = s_j,$$

where

$$s_j = \begin{cases} 1 \text{ if } (\theta_{nj} - \theta_{0n,j}) \geq 0 \text{ and } j \in A_0 \\ -1 \text{ if } (\theta_{nj} - \theta_{0n,j}) < 0 \text{ and } j \in A_0 \\ 0 \text{ if} j \in \bar{A}_0. \end{cases}$$

Set $s = (s_1, ..., s_{p_n})'$. If $\left\| \theta_{nA_0} - \theta_{0nA_0} \right\|_1 = \tau_n$ and $j \in A_0$, then $\tilde{p}_{\lambda_n}(\theta_{nj}) - \tilde{p}_{\lambda_n}(\theta_{0n,j}) = 0$. If, in addition, $\theta_{n\bar{A}_0} = 0$, the KKT conditions for (B1.4) are

(B1.5) $\dfrac{\partial Q_{n\infty}(\theta_n)}{\partial \theta_{nj}} + \nu s_j = 0$

if $j \in A_0$, where $\nu$ is a Lagrangian multiplier, and

(B1.6) $\left| \dfrac{\partial Q_{n\infty}(\theta_n)}{\partial \theta_{nj}} \right| \leq \lambda_n$

if $j \in \bar{A}_0$. By a Taylor series expansion,

$$\frac{\partial Q_{n\infty}(\theta_n)}{\partial \theta} = \frac{\partial Q_{n\infty}(\breve{\theta}_n)}{\partial \theta \, \partial \theta'}(\theta_n - \theta_{0n})$$

where $\breve{\theta}_n$ is the Taylor series intermediate point and may be different in different occurrences. Therefore, (B1.5) and (B1.6) can be written as

(B1.7) $H_{n11}(\breve{\theta}_n)(\theta_{nA_0} - \theta_{0nA_0}) + \nu s_{A_0} = 0$

(B1.8) $\left| H_{n21}(\breve{\theta}_n)(\theta_{nA_0} - \theta_{0nA_0}) \right| \leq \lambda_n$,

where $s_{A_0} = (s_1, ..., s_{p_0})'$. By (B1.7) and assumption 4(i),

$$(\theta_{nA_0} - \theta_{0nA_0}) = -\nu H_{n11}^{-1}(\breve{\theta}_n) s_{A_0},$$

so $\left\| \theta_{nA_0} - \theta_{0nA_0} \right\|_1 = \tau_n$ implies that

$$|\nu|\sum_{j=1}^{p_0}\left|\left(H_{n11}^{-1}(\breve{\theta}_n)s_{A_0}\right)_j\right| = \tau_n .$$

Define

$$C_{nA_0} = \sum_{j=1}^{p_0}\left|\left(H_{n11}^{-1}(\breve{\theta}_n)s_{A_0}\right)_j\right| .$$

Then

$$|\nu| = \tau_n / C_{nA_0}$$

and

$$|(\theta_{nA_0} - \theta_{0nA_0})_j| = \frac{\tau_n}{C_{nA_0}}\left|(H_{n11}^{-1}(\breve{\theta}_n)s_{A_0})_j\right|$$

for each $j = 1,..., p_0$. Inequality (B1.8) is

$$C_{nA_0}^{-1}\left|H_{n21}(\breve{\theta}_n)H_{n11}^{-1}(\breve{\theta}_n)s_{A_0}\right| \le \lambda_n / \tau_n .$$

By assumption 4(ii), this holds for all sufficiently large $n$ because $\tau_n \ll \lambda_n$ and $C_{nA_0}$ is bounded away from 0 for all $n$. It follows that the KKT conditions are satisfied. Q.E.D.

Proof of Theorem 2.1: The proof consists of proving equations (2.5) and (2.6).

*Equation* (2.5): $S_n(\chi_n, \hat{\theta}_n) - S_{n\infty}(\hat{\theta}_n) = Q_n(\chi_n, \hat{\theta}_n) - Q_{n\infty}(\hat{\theta}_n)$. Equation (2.5) now follows from assumption 3(i) and lemmas B.2 and B.3.

*Equation* (2.6): $P\left[T_n(\hat{\theta}_{n\hat{A}_0}^{PO}) > \hat{c}_\alpha^{PO} \mid \hat{A}_0 = A_0\right] = P\left[T_n(\hat{\theta}_{nA_0}^{O}) > \hat{c}_\alpha^{O}\right] = \alpha$ by the definition of $\hat{c}_\alpha^{PO}$. By (2.5),

$$P\left[T_n(\hat{\theta}^{PO}_{n\hat{A}_0}) > \hat{c}^{PO}_\alpha\right] = P\left[T_n(\hat{\theta}^{PO}_{n\hat{A}_0}) > \hat{c}^{PO}_\alpha \mid \hat{A}_0 = A_0\right] P(\hat{A}_0 = A_0)$$

$$+ P\left[T_n(\hat{\theta}^{PO}_{n\hat{A}_0}) > \hat{c}^{PO}_\alpha \mid \hat{A}_0 \neq A_0\right] P(\hat{A}_0 \neq A_0)$$

$$= P\left[T_n(\hat{\theta}^{PO}_{n\hat{A}_0}) > \hat{c}^{PO}_\alpha \mid \hat{A}_0 = A_0\right]$$

$$+ \left\{P\left[T_n(\hat{\theta}^{PO}_{n\hat{A}_0}) > \hat{c}^{PO}_\alpha \mid \hat{A}_0 \neq A_0\right] - P\left[T_n(\hat{\theta}^{PO}_{n\hat{A}_0}) > \hat{c}^{PO}_\alpha \mid \hat{A}_0 = A_0\right]\right\} P(\hat{A}_0 \neq A_0)$$

$$= P\left[T_n(\hat{\theta}^{PO}_{n\hat{A}_0}) > \hat{c}^{PO}_\alpha \mid \hat{A}_0 = A_0\right] + o(n^{-2})$$

$$= \alpha + o(n^{-2}).$$

Therefore,

$$\left| P\left[T_n(\hat{\theta}^{PO}_{n\hat{A}_0}) > \hat{c}^{PO}_\alpha\right] - P\left[T_n(\hat{\theta}^{O}_{nA_0}) > c^{O}_\alpha\right] \right| = o(n^{-2}). \text{ Q.E.D.}$$

## B.2 PROOF OF THEOREM *2.2*

Define the population pseudo-true value

$$\breve{\theta}_{0nA_0} = \arg \min_{(\theta_{nA_0}, 0_{p_n - p_0}) \in \Theta_n} Q_{n\infty}(\theta_{nA_0}, 0_{p_n - p_0})$$

and $\breve{\theta}_{0n} = (\breve{\theta}'_{0nA_o}, 0'_{p_n - p_0})'$

Lemma B.5: Let assumptions 1-4, 1S and 2S of Appendix A hold. Then

(i) $\left\| \breve{\theta}_{0nA_0} - \theta_{0nA_0} \right\|_1 = o(\tau_n)$.

(ii) $S_{n\infty}(\theta_n)$ is minimized at $(\breve{\theta}_{0nA_0}, 0_{0n\bar{A}_0})$.

(iii) $\hat{\theta}_n - \breve{\theta}_{0n} \to^p 0$ as $n \to \infty$, and $\lim_{n\to\infty} P(\hat{A}_0 = A_0) = 1$.

Proof:

Part (i): By a Taylor series expansion,

$$Q_n(\theta_n) = Q_n(\theta_{0n}) + (1/2)(\theta_n - \theta_{0n})' H_n(\breve{\theta}_n)(\theta_n - \theta_{0n}),$$

where $H_n(\theta_n) = \partial^2 Q_n(\theta_n) / \partial\theta_n \partial\theta_n'$ and $\breve{\theta}_n$ is the Taylor series intermediate point. Define $\breve{H}_n = H_n(\breve{\theta}_n)$. Then

$$\begin{aligned}&(\theta_n - \theta_{0n})' H_n(\breve{\theta}_n)(\theta_n - \theta_{0n}) \\ &\quad = \left[(\theta_{nA_0} - \theta_{0nA_0})' \quad (\theta_{n\bar{A}_0} - \theta_{0n\bar{A}_0})'\right] \begin{pmatrix} \breve{H}_{n11} & \breve{H}_{n12} \\ \breve{H}_{n21} & \breve{H}_{n22} \end{pmatrix} \begin{bmatrix} (\theta_{nA_0} - \theta_{0nA_0}) \\ (\theta_{n\bar{A}_0} - \theta_{0n\bar{A}_0}) \end{bmatrix}\end{aligned}$$

and

$$\frac{\partial Q_{n\infty}(\theta_n)}{\partial \theta_{nA_0}} = \breve{H}_{n11}(\theta_{nA_0} - \theta_{0nA_0}) + \breve{H}_{n12}(\theta_{n\bar{A}_0} - \theta_{0n\bar{A}_0}).$$

By the definition of $\breve{\theta}_{0nA_0}$

$$\frac{\partial Q_{n\infty}(\breve{\theta}_{0nA_0}, 0_{p_n - p_0})}{\partial \theta_{nA_0}} = \breve{H}_{n11}(\breve{\theta}_{0nA_0} - \theta_{0nA_0}) - \breve{H}_{n12}\theta_{0n\bar{A}_0} = 0.$$

Therefore,

$$(\breve{\theta}_{0nA_0} - \theta_{0nA_0}) = \breve{H}_{n11}^{-1}\breve{H}_{n12}\theta_{0n\bar{A}_0},$$

and by assumption 1S,

(B.2.1) $\left|(\breve{\theta}_{0nA_0} - \theta_{0nA_0})_j\right| = o(\tau_n)$.

Part (ii): (B.2.1) implies that $|\breve{\theta}_{0nA_0,j}| \gg \lambda_n$. Therefore, replacing $\theta_{0n}$ with $\breve{\theta}_{0n} = (\breve{\theta}_{0nA_0}', 0_{p_n - p_0}')'$ in the proof of Lemma B.1 and using assumption 1S yields Part (ii).

Part (iii): The conclusions and proofs of Lemmas B.2 and B.3 remain unchanged after replacing $\theta_{0n}$ with $\breve{\theta}_{0n} = (\breve{\theta}_{0nA_0}, 0_{p_n - p_0})$. Therefore, it follows from Assumption 3(i) that $\hat{\theta}_n - \breve{\theta}_{0n} \to^p 0$ as $n \to \infty$ and $P(\hat{A}_0 = A_0) \to 1$. Q.E.D.

Part (i) of Lemma B.5 shows that the true and pseudo-true parameter values differ by $o(\tau_n)$. Part (ii) shows that the penalization and thresholding procedure drives the small non-zero parameters to zero and replaces the true values of the large parameters with the pseudo-true values $\breve{\theta}_{0nA_0}$. Part(iii) shows that the penalization and thresholding procedure estimates the parameters of the pseudo-true model consistently and discriminates correctly between large and small parameters as $n \to \infty$.

<u>Proof of Theorem 2,2</u>: By assumption 3S and part (iii) of Lemma B.4, it suffices to prove that

$$\sup_z \left| P^* \left( \frac{\theta^{*PT}_{nA_0,j} - \hat{\theta}^{PT}_{nA_0,j}}{s^{*PT}_{A_0}} \le z \right) - P \left( \frac{\hat{\theta}^{PT}_{nA_0,j} - \theta_{0nA_0,j}}{s^{PT}} \le z \right) \right| \to^p 0$$

as $n \to \infty$, where $s^{*PT}_{A_0}$ is the bootstrap standard error obtained when $\hat{A}_0$ is replaced with $A_0$. Now

$$P \left( \frac{\hat{\theta}^{PT}_{nA_0,j} - \theta_{0nA_0,j}}{s^{PT}} \le z \right) = P \left( \frac{\hat{\theta}^{PT}_{nA_0,j} - \breve{\theta}_{0nA_0,j}}{s^{PT}} + \frac{\breve{\theta}_{0nA_0,j} - \theta_{0nA_0,j}}{s^{PT}} \le z \right)$$

$$= P \left( \frac{\hat{\theta}^{PT}_{nA_0,j} - \breve{\theta}_{0nA_0,j}}{s^{PT}} \le z \right) + o(1)$$

uniformly over $-\infty < z < \infty$. Therefore,

$$\sup_z \left| P^*\left( \frac{\theta^{*PT}_{nA_0,j} - \hat{\theta}^{PT}_{nA_0,j}}{s^{*PT}_{A_0}} \le z \right) - P\left( \frac{\hat{\theta}^{PT}_{nA_0,j} - \theta_{0nA_0,j}}{s^{PT}} \le z \right) \right|$$

$$= \sup_z \left| P^*\left( \frac{\theta^{*PT}_{nA_0,j} - \hat{\theta}^{PT}_{nA_0,j}}{s^{*PT}_{A_0}} \le z \right) - P\left( \frac{\hat{\theta}^{PT}_{nA_0,j} - \breve{\theta}_{0nA_0,j}}{s^{PT}} \le z \right) \right| + o(1),$$

and the result follows from Assumption 3S. Q.E.D.

## B.3 CONDITIONS SUFFICIENT FOR SATISFACTION OF ASSUMPTION 3(i)

Proposition 1: For each $\theta \in [0,1]^{p_n}$, let $g(X,\theta)$ be a measurable function of the possibly vector valued random variable $X$. Assume:

(i) $\{X_i : i = 1,...,n\}$ is an independent random sample from the distribution of $X$.

(ii) $\theta \in \Theta_n = [0,1]^{p_n}$, $\|\theta\|_1 \le C$ for some $C < \infty$, every $n$, and $p_n = n^b$ for some $b < 1$.

(iii) $E[g(X,\theta)] = 0$ and $E[g(X,\theta)^2] \le \sigma_g^2$ for some constant $\sigma_g^2 < \infty$, all $\theta \in \Theta_n$, and all $n$.

(iv) There is a constant $K_g < \infty$ not depending on $n$ such that $E|g(X,\theta)|^\ell \le \ell!\sigma_g^2 K_g^{\ell-2}$ $(\ell = 3,4,...)$ for all $\theta \in \Theta_n$ and all $n$.

(v) For each $n$, there is a function $M_n(X)$ such that

$$|g(X,\theta_1) - g(X,\theta_2)| \le M_n(X)\|\theta_1 - \theta_2\|_1$$

for all $\theta_1, \theta_2 \in \Theta_n$. Moreover, there are finite constants $M^*$ and $K_M$ not depending on $n$ such that $|EM_n(X)| \le M^*$ and $E|M_n(X) - EM_n(X)|^\ell \le \ell!\sigma_{M_n}^2 K_M^{\ell-2}$ ($\ell = 3,4...$) for each $n$, where $\sigma_{M_n}^2 = Var[M_n(X)]$ and $0 < m_0 \le \sigma_{M_n}^2 \le M_0$ for some finite $M_0, m_0 > 0$.

Then, there are finite constants $c > 0$, $\varepsilon_0 > 0$, $n_0 > 0$, and $\ell > 0$ such that

$$P\left[\sup_{\theta\in\Theta_n}\left|n^{-1}\sum_{i=1}^{n} g(X_i,\theta)\right| > \varepsilon\right] \le 3m_n \exp(-cn\varepsilon^2)$$

for all $n > n_0$ if $\ell\tau_n^2 \le \varepsilon < \varepsilon_0$.

Proof: Define $m_n$ as in Appendix A. Divide $\Theta_n$ into $m_n$ hypercubic cells whose edges have lengths $d_n = m_n^{-1/p_n}$. Denote the cells by $\Theta_{nj}$ ($j = 1,...,m_n$). Let $\theta^j$ be a point in the interior of $\Theta_{nj}$. Let $\varepsilon > 0$ be arbitrary. Define

$$P_n = P\left[\sup_{\theta\in\Theta_n}\left|n^{-1}\sum_{i=1}^{n} g(X_i,\theta)\right| > \varepsilon\right].$$

Then

$$P_n = P\left\{\max_{1\le j\le m_n}\left[\sup_{\theta\in\Theta_{nj}}\left|n^{-1}\sum_{i=1}^{n} g(X_i,\theta)\right| > \varepsilon\right]\right\}$$

$$= P\left\{\max_{1\le j\le m_n}\left[\sup_{\theta\in\Theta_{nj}}\left|n^{-1}\sum_{i=1}^{n} g(X_i,\theta^j) + n^{-1}\sum_{i=1}^{n}[g(X_i,\theta) - g(X,\theta^j)]\right| > \varepsilon\right]\right\}$$

$$\le P\left\{\max_{1\le j\le m_n}\left[\left|n^{-1}\sum_{i=1}^{n} g(X_i,\theta^j)\right| + \sup_{\theta\in\Theta_{nj}}\left|n^{-1}\sum_{i=1}^{n}[g(X_i,\theta) - g(X,\theta^j)]\right| > \varepsilon\right]\right\}$$

$$\le P\left[\max_{1\le j\le m_n}\left|n^{-1}\sum_{i=1}^{n} g(X_i,\theta^j)\right| > \varepsilon/2\right] + P\left\{\max_{1\le j\le m_n}\sup_{\theta\in\Theta_{nj}}\left|n^{-1}\sum_{i=1}^{n}[g(X_i,\theta) - g(X,\theta^j)]\right| > \varepsilon/2\right\}$$

$$\equiv P_{n1} + P_{n2}.$$

Consider $P_{n1}$. By Bernstein's inequality

$$P_{n1} = P\left[\max_{1\le j\le m_n}\left|n^{-1}\sum_{i=1}^{n} g(X_i,\theta^j)\right| > \varepsilon/2\right] \le 2m_n \exp(-n\varepsilon^2/32\sigma_g^2).$$

if $\varepsilon < 4\sigma_g^2 / K_g$.

Now consider $P_{n2}$. By assumption (v)

$$P_{n2} = P\left\{ \max_{1\le j\le m_n} \sup_{\theta\in\Theta_{nj}} \left| n^{-1}\sum_{i=1}^{n}[g(X_i,\theta) - g(X,\theta^j)] \right| > \varepsilon / 2 \right\}$$

$$\le P\left\{ \max_{1\le j\le m_n} \sup_{\theta\in\Theta_{nj}} n^{-1}\sum_{i=1}^{n} | g(X_i,\theta) - g(X,\theta^j) | > \varepsilon / 2 \right\}$$

$$\le P\left\{ \max_{1\le j\le m_n} \sup_{\theta\in\Theta_{nj}} n^{-1}\sum_{i=1}^{n} M_n(X_i) \left\|\theta - \theta^j\right\|_1 > \varepsilon / 2 \right\}$$

$$\le m_n P\left[ d_n p_n n^{-1}\sum_{i=1}^{n} M_n(X_i) > \varepsilon / 2 \right]$$

$$= m_n P\left[ n^{-1}\sum_{i=1}^{n} M_n(X_i) > \frac{\varepsilon}{2 d_n p_n} \right].$$

Therefore,

$$P_{n2} \le m_n P\left[ n^{-1}\sum_{i=1}^{n} [M_n(X_i) - EM_n(X)] > \frac{\varepsilon}{2 d_n p_n} - EM_n(X) \right]$$

$$\le m_n P\left[ n^{-1}\sum_{i=1}^{n} [M_n(X_i) - EM_n(X)] > \frac{\varepsilon}{2 d_n p_n} - M^* \right].$$

By Bernstein's inequality,

$$P_{n2} \le 2m_n \exp\left\{-\frac{n[\varepsilon/(2d_n p_n) - M^*]^2}{4M_0 + 2K_M[\varepsilon/(2d_n p_n) - M^*]}\right\}$$

$$\le 2m_n \exp\left\{-\frac{n[\varepsilon/(4d_n p_n)]^2}{4M_0 + 2K_M\varepsilon/(d_n p_n)}\right\}$$

if $\varepsilon > 4d_n p_n M^*$. Now let

$$\varepsilon > 4d_n p_n \max(M^*, M_0/K_M).$$

Then

$$P_{n2} \le 2m_n \exp\left[-\frac{n}{32K_M}\frac{\varepsilon}{d_n p_n}\right].$$

Let $n_0$ be the smallest value of $n$ such that

$$[4d_n p_n \max(M^*, M_0/K_M)]^{1/2} < \ell\tau_n^2 < \frac{\sigma_g^2}{K_M d_n p_n}$$

for some $\ell > 0$. Then if $n > n_0$ and $\ell\tau_n^2 \le \varepsilon < \sigma_g^2/(K_M d_{n_0} p_{n_0})$,

$$P_{n2} \le 2m_n \exp\left(-\frac{n\varepsilon^2}{32\sigma_g^2}\right).$$

Set

$$\varepsilon_0 = \min\left(\frac{4\sigma_g^2}{K_g}, \frac{\sigma_g^2}{K_M d_{n_0} p_{n_0}}\right)$$

and $c = 1/(32\sigma_g^2)$. Q.E.D.

B.3.1 *Examples of Models that Satisfy Assumption 3(i)*

Example 1: Penalized least squares estimation of a linear model. The model is

(B3.1.1) $$Y_i = \sum_{j=1}^{p_n} \theta'_{0n,j} X_{ij} + U_i \equiv \theta'_{0n} X_i + U_i; \quad i = 1,...,n,$$

where the $X_{ij}$'s are random variables; $\|X_i\|_1 \leq M$ for some $M < \infty$, all $i = 1,2,...$; and all $j = 1,..., p_n$; the $U_i$'s are independently and identically distributed sub-Gaussian random variables; $E(U_i) = E(U_i \mid X_i) = 0$; and $E(U_i^2) = \sigma^2$ for all $i = 1,...,n$ and $j = 1,..., p_n$. Assume that $\|\theta\|_1 \leq M$ for all $\theta \in \Theta_n$ and all $n$. Also assume that $p_n = O(n^b)$ for some $0 \leq b < 1$. Let

$$Q_n(\chi_n, \theta) = n^{-1} \sum_{i=1}^{n} (Y_i - \theta' X_i)^2 = n^{-1} \sum_{i=1}^{n} [U_i - (\theta - \theta_{0n})' X_i]^2,$$

and

$$Q_{n\infty}(\theta) = EQ_n(\chi_n, \theta) = \sigma^2 + (\theta - \theta_{on})' \Sigma_{XX} (\theta - \theta_{0n}),$$

where $\Sigma_{XX} = E(XX')$. In the notation of Proposition 1,

$$g(X, \theta) \to g(X, U, \theta) = (U^2 - \sigma^2) - 2U(\theta - \theta_{0n})' X + (\theta - \theta_{0n})'(XX' - \Sigma_{XX})(\theta - \theta_{0n}).$$

We show that model (B3.1.1) satisfies the conditions of Proposition 1.

Conditions (i), (ii), and (iii) are satisfied by the definition of the model and by the arguments below for condition (iv) with $\ell = 2$.

Condition (iv): Let $X_{\bullet j}$ denote the $j$'th component of $X$. Then,

$$|g(X, U, \theta)| \leq |U^2 - \sigma^2| + 2|U| |(\theta - \theta_{0n})' X| + |(\theta - \theta_{0n})'(XX' - \Sigma_{XX})(\theta - \theta_{0n})|.$$

$$|(\theta - \theta_{0n})' X| = \left| \sum_{j=1}^{p_n} (\theta - \theta_{0n})_j X_{\bullet j} \right| \leq \sum_{j=1}^{p_n} |(\theta - \theta_{0n})_j| |X_{\bullet j}| \leq M \|\theta - \theta_{0n}\|_1 \leq 2M^2$$

$$(\theta - \theta_{0n})' XX' (\theta - \theta_{0n}) \leq 4M^4$$

$$(\theta - \theta_{0n})' \Sigma_{XX} (\theta - \theta_{0n}) \leq 4M^4$$

$$|g(u, X, \theta| \leq |U^2 - \sigma^2| + 4|U|M^2 + 8M^4$$

Therefore,

$$E|g(U, X, \theta)|^{\ell} \leq E\left||U^2 - \sigma^2| + 4|U|M^2 + 8M^4\right|^{\ell}$$

$$\leq 2^{\ell} E|U^2 - \sigma^2|^{\ell} + 2^{\ell} E(4M^2|U| + 8M^4)^{\ell}$$

$$\leq 2^{\ell} E\left|U^2 - \sigma^2\right|^{\ell} + 2^{4\ell} M^{2\ell} E|U|^{\ell} + 2^{2\ell}(8M^4)^{\ell}.$$

The first and second terms on the right-hand side of the inequality satisfy condition (iv) because $U$ is sub-Gaussian and $U^2 - \sigma^2$ is sub-exponential. The third term satisfies condition (iv) because it is a constant.

Condition (v): $g(U, X, \theta)$ is continuously differentiable with respect to $\theta$. Therefore,

$$|g(U, X, \theta_2) - g(U, X, \theta_1)| \leq \left|\frac{\partial g(U, X, \tilde{\theta})}{\partial \theta'}(\theta_2 - \theta_1)\right|,$$

where $\tilde{\theta}$ is the Taylor series intermediate point.

$$\frac{\partial g(U, X, \tilde{\theta})}{\partial \theta} = -2[X'U - (\tilde{\theta} - \theta_{0n})'(XX' - \Sigma_{XX})]$$

$$\left|\frac{\partial g(U, X, \tilde{\theta})}{\partial \theta_j}(\theta_2 - \theta_1)\right| \leq 2M^2|U| + 8M^4.$$

Condition (v) is satisfied because $U$ is sub-Gaussian.

Example 2: A Binary logit model with normally distributed random coefficients. Let $\{Y_i, X_i : i = 1, ..., n\}$ be independently and identically distributed realizations of the binary random variable $Y \in \{0,1\}$ and the $d_X \times 1$ random vector $X$. Let $X_{ij}$ denote the $j$'th component of $X_i$, and assume that $\|X\|_1 \le M$ for all $(i, j)$ and some $M < \infty$ Define $\theta = vec(\beta, C)$, where $C$ is a $d_X \times d_X$ Cholesky factorization matrix. The model is

$$\text{(B3.1.2)} \qquad P(Y_i = 1 \mid X_i, \theta) = \int \left\{ \frac{\exp[(\beta' + \varepsilon' C') X_i]}{1 + \exp[(\beta' + \varepsilon' C') X} \right\} f(\varepsilon) d\varepsilon ,$$

where $f$ is the $N(0, I_{d_X})$ probability density function. In the notation of Proposition 1, $g(x, \theta) = P(Y_i = 1 \mid X = x, \theta)$. Assume that $g(x, \theta)$ is bounded away from 0 and 1 for all $x \in \text{supp}(X)$ and all $\theta \in \Theta_n$. The log-likelihood function for estimating $\beta$ and $C$ is

$$Q_n(\chi_n, \theta) = n^{-1} \sum_{i=1}^{n} \{ Y_i \log g(X_i, \theta) + (1 - Y_i) \log[1 - g(X_i, \theta)] \} ,$$

Let

$$\begin{aligned} Q_{n\infty}(\theta) &= E[P(Y = 1 \mid X) \log g(X, \theta)] + E\{P(Y = 0 \mid X) \log[1 - g(X_i, \theta)]\} \\ &= E[g(X, \theta_{0n}) \log g(X, \theta)] + E\{[1 - g(X, \theta_{0n})] \log[1 - g(X, \theta)]\}, \end{aligned}$$

where $\theta_{0n} \in \Theta_n$. We show that model (B.3.3.2) satisfies the conditions of Proposition 1.

Conditions (i) and (ii) are satisfied by the definition of the model.

Conditions (iii) and (iv) are satisfied because $| Y_i \log g(X_i, \theta) + (1 - Y_i) \log[1 - g(X_i, \theta)] |$ is bounded for all $i = 1, ..., n$.

Condition (v): This condition is satisfied if $| g(X, \theta_2) - g(X, \theta_1) | \le M_g \|\theta_2 - \theta_1\|_1$, where $M_g < \infty$ is a constant. To establish this inequality, define

$$\pi(X, \varepsilon, \beta, C) = \frac{\exp[(\beta' + \varepsilon' C') X]}{1 + \exp[(\beta' + \varepsilon' C') X} .$$

Then

$$\frac{\partial \pi}{\partial \beta_j} = X_j \pi(1-\pi)$$

and

$$\frac{\partial \pi}{\partial C_{jk}} = \varepsilon_j X_k \pi(1-\pi) .$$

A Taylor series expansion gives

$$| g(X,\theta_2) - g(X,\theta_1) | \le \int | \pi(X,\varepsilon,\beta_2,C_2) - \pi(X,\varepsilon,\beta_1,C_1) | f(\varepsilon) d\varepsilon$$

$$= \int \sum_{j=1}^{\dim(\theta)} \left| \frac{\partial \pi(X,\varepsilon,\tilde{\beta},\tilde{C})}{\partial \theta_j} \right| | \theta_{1j} - \theta_{2j} | f(\varepsilon) d\varepsilon$$

$$= \sum_{j=1}^{\dim(\theta)} | \theta_{1j} - \theta_{2j} | \int \left| \frac{\partial \pi(X,\varepsilon,\tilde{\beta},\tilde{C})}{\partial \theta_j} \right| f(\varepsilon) d\varepsilon,$$

where $\tilde{\beta}$ and $\tilde{C}$, respectively, are a vector and matrix of Taylor series intermediate points. Let $\mu = (2/\pi)^{1/2}$ denote the first absolute moment of the $N(0,1)$ distribution. Because $\pi(1-\pi) \le 0.25$ and $| X_{ij} | \le M$,

$$| g(X,\theta_2) - g(X,\theta_1) | \le 0.25 M \int \left( \sum_{j=1}^{d_X} | \beta_{1j} - \beta_{2j} | + \sum_{j,k=1}^{d_X} | \varepsilon_j || C_{2jk} - C_{1jk} | \right) f(\varepsilon) d\varepsilon$$

$$\le 0.25 M \left\| \beta_2 - \beta_1 \right\|_1 + 0.25 M \mu \sum_{j,k=1}^{d_X} | C_{2jk} - C_{1jk} |$$

$$\le 0.25 M \left\| \beta_2 - \beta_1 \right\|_1 + 0.25 M \sum_{j,k=1}^{d_X} | C_{2jk} - C_{1jk} |.$$

Therefore,

$$| g(X,\theta_2) - g(X,\theta_1) | \le 0.25 M \left\| \theta_2 - \theta_1 \right\|_1 .$$

Set $M_g = 0.25M$ .

Example 3: A Generalized Method of Moments Estimator with a Fixed Weight Matrix.

Let

$$Q_n(\chi_n,\theta) = \left[ n^{-1}\sum_{i=1}^{n} g(X_i,\theta) \right]' \Omega \left[ n^{-1}\sum_{i=1}^{n} g(X_i,\theta) \right]$$

and

$$Q_{n\infty}(\theta) = [Eg(X,\theta)]'\Omega[Eg(X,\theta)],$$

where $g$ is a $q\times 1$ vector-valued function and $\Omega$ is a $q\times q$ positive definite symmetrical matrix of finite constants. Assume each component of $g$ satisfies the conditions of Proposition 1. Then Assumption 3(i) is satisfied.

Example 4: A Generalized Method of Moments Estimator with a the Continuous Updating estimate of the Asymptotically Optimal Weight Matrix.

Use the notation of Example 3 and assume that $\Omega_0(\theta) \equiv \text{cov}[g(X,\theta)g(X,\theta)']$ is positive definite with bounded eigenvalues for all $\theta \in \Theta_n$ and $n$. Also assume that the components of $\Omega_0(\theta)$ satisfy $|\Omega_{0,ij}(\theta)| \le M$ for some $M < \infty$ and all $\theta \in \Theta_n$, $i,j$, and $n$. Then $\Omega_0(\theta)$

$$Q_n(\chi_n,\theta) = \left[ n^{-1}\sum_{i=1}^{n} g(X_i,\theta) \right]' \Omega_n^{-1}(\theta) \left[ n^{-1}\sum_{i=1}^{n} g(X_i,\theta) \right]$$

where

$$\Omega_n(\theta) = n^{-1}\sum_{i=1}^{n} g(X_i,\theta)g(X_i,\theta)' - \overline{g}_n(\theta)\overline{g}_n(\theta),$$

$$\overline{g}_n(\theta) = n^{-1}\sum_{i=1}^{n} g(X_i,\theta),$$

and

$$Q_{n\infty}(\theta) = [Eg(X,\theta)]'\Omega_0^{-1}(\theta)[Eg(X,\theta)].$$

Assumption 3(i) is satisfied if each component of $g$ satisfies the conditions of Proposition 1.

## B.4 LOGIT CONFIDENCE INTERVALS

This section presents confidence intervals for the logit model of Section ?? averaged over Monte Carlo replications. Confidence intervals for the full model are not shown because, as Tables 1-6 show, the differences between their true and nominal coverage probabilities are very large in most cases.

Table B.4.1: Logit Confidence Intervals for $\theta_{0n,1}$: Nominal level 0.90 $p_n = n/10$

| $n$ | Interval | Oracle Model Asymp. | Oracle Model Boot. | Pseudo Oracle Model Asymp., $C_n = 1$ | Pseudo Oracle Model Boot., $C_n = 1$ | Pseudo Oracle Model Asymp. $C_n = \log\log p_n$ | Pseudo Oracle Model Boot., $C_n = \log\log p_n$ |
|---|---|---|---|---|---|---|---|
| 500 | Lower 1-Sided | $(-\infty, 5.74)$ | $(-\infty, 5.02)$ | $(-\infty, 5.72)$ | $(-\infty, 5.01)$ | $(-\infty, 5.54)$ | $(-\infty, 4.87)$ |
| 1000 | | $(-\infty, 4.88)$ | $(-\infty, 4.61)$ | $(-\infty, 4.91)$ | $(-\infty, 4.63)$ | $(-\infty, 4.88)$ | $(-\infty, 4.61)$ |
| 2000 | | $(-\infty, 4.52)$ | $(-\infty, 4.40)$ | $(-\infty, 4.53)$ | $(-\infty, 4.40)$ | $(-\infty, 4.52)$ | $(-\infty, 4.40)$ |
| 4000 | | $(-\infty, 4.34)$ | $(-\infty, 4.28)$ | $(-\infty, 4.34)$ | $(-\infty, 4.28)$ | $(-\infty, 4.34)$ | $(-\infty, 4.29)$ |
| | | | | | | | |
| 500 | Upper 1-Sided | $(3.79, \infty)$ | $(3.29, \infty)$ | $(3.78, \infty)$ | $(3.28, \infty)$ | $(3.67, \infty)$ | $(3.20, \infty)$ |
| 1000 | | $(3.73, \infty)$ | $(3.48, \infty)$ | $(3.75, \infty)$ | $(3.50, \infty)$ | $(3.73, \infty)$ | $(3.48, \infty)$ |
| 2000 | | $(3.75, \infty)$ | $(3.64, \infty)$ | $(3.76, \infty)$ | $(3.64, \infty)$ | $(3.75, \infty)$ | $(3.64, \infty)$ |
| 4000 | | $(3.81, \infty)$ | $(3.75, \infty)$ | $(3.81, \infty)$ | $(3.75, \infty)$ | $(3.81, \infty)$ | $(3.75, \infty)$ |
| | | | | | | | |
| 500 | Symmetrical | (3.51,6.01) | (3.27,6.25) | (3.50,6.00) | (3.27,6.23) | (3.40,5.80) | (3.18,6.02) |
| 1000 | | (3.56,5.05) | (3.46,5.15) | (3.58,5.08) | (3.48,5.18) | (3.56,5.05) | (3.46,5.15) |
| 2000 | | (3.64,4.63) | (3.61,4.66) | (3.65,4.64) | (3.61,4.67) | (3.64,4.63) | (3.61,4.66) |
| 4000 | | (3.73,4.41) | (3.72,4.42) | (3.73,4.41) | (3.72,4.43) | (3.73,4.41) | (3.72,4.42) |

Table B.4.2: Logit Confidence Intervals for $\theta_{0n,1}$: Nominal level 0.90 $p_n = n/2$

| $n$ | Interval | Oracle Model Asymp. | Oracle Model Boot. | Pseudo Oracle Model Asymp., $C_n = 1$ | Pseudo Oracle Model Boot., $C_n = 1$ | Pseudo Oracle Model Asymp. $C_n = \log\log p_n$ | Pseudo Oracle Model Boot., $C_n = \log\log p_n$ |
|---|---|---|---|---|---|---|---|
| 500 | Lower 1-Sided | $(-\infty, 5.75)$ | $(-\infty, 5.04)$ | $(-\infty, 6.06)$ | $(-\infty, 5.25)$ | $(-\infty, 5.15)$ | $(-\infty, 4.57)$ |
| 1000 | | $(-\infty, 4.89)$ | $(-\infty, 4.61)$ | $(-\infty, 4.95)$ | $(-\infty, 4.66)$ | $(-\infty, 4.88)$ | $(-\infty, 4.60)$ |
| 2000 | | $(-\infty, 4.52)$ | $(-\infty, 4.40)$ | $(-\infty, 4.55)$ | $(-\infty, 4.42)$ | $(-\infty, 4.52)$ | $(-\infty, 4.40)$ |
| 4000 | | $(-\infty, 4.35)$ | $(-\infty, 4.29)$ | $(-\infty, 4.36)$ | $(-\infty, 4.39)$ | $(-\infty, 4.35)$ | $(-\infty, 4.29)$ |
| | | | | | | | |
| 500 | Upper 1-Sided | $(3.80, \infty)$ | $(3.29, \infty)$ | $(3.95, \infty)$ | $(3.41, \infty)$ | $(3.44, \infty)$ | $(3.02, \infty)$ |
| 1000 | | $(3.73, \infty)$ | $(3.48, \infty)$ | $(3.77, \infty)$ | $(3.51, \infty)$ | $(3.72, \infty)$ | $(3.48, \infty)$ |
| 2000 | | $(3.75, \infty)$ | $(3.64, \infty)$ | $(3.77, \infty)$ | $(3.65, \infty)$ | $(3.75, \infty)$ | $(3.64, \infty)$ |
| 4000 | | $(3.82, \infty)$ | $(3.76, \infty)$ | $(3.83, \infty)$ | $(3.77, \infty)$ | $(3.82, \infty)$ | $(3.76, \infty)$ |
| | | | | | | | |
| 500 | Symmetrical | (3.52,6.03) | (3.28,6.28) | (3.66,6.35) | (3.40,6.61) | (3.20,5.40) | (3.00,5.59) |
| 1000 | | (3.56,5.05) | (3.46,5.15) | (3.60,5.11) | (3.49,5.22) | (3.56,5.05) | (3.46,5.15) |
| 2000 | | (3.64,4.63) | (3.61,4.17) | (3.66,4.65) | (3.63,4.69) | (3.64,4.63) | (3.61,4.67) |
| 4000 | | (3.74,4.43) | (3.73,4.44) | (3.75,4.44) | (3.72,4.43) | (3.74,4.45) | (3.73,4.44) |

Table B.4.3: Logit Confidence Intervals for $\theta_{0n,1}$: Nominal level 0.90 $p_n = 3n/4$

| $n$ | Interval | Oracle Model Asymp. | Oracle Model Boot. | Pseudo Oracle Model Asymp., $C_n = 1$ | Pseudo Oracle Model Boot., $C_n = 1$ | Pseudo Oracle Model Asymp. $C_n = \log\log p_n$ | Pseudo Oracle Model Boot., $C_n = \log\log p_n$ |
|---|---|---|---|---|---|---|---|
| 500 | Lower 1-Sided | $(-\infty, 5.81)$ | $(-\infty, 5.06)$ | $(-\infty, 6.39)$ | $(-\infty, 5.48)$ | $(-\infty, 5.17)$ | $(-\infty, 4.57)$ |
| 1000 | | $(-\infty, 4.89)$ | $(-\infty, 4.61)$ | $(-\infty, 5.03)$ | $(-\infty, 4.73)$ | $(-\infty, 4.89)$ | $(-\infty, 4.61)$ |
| 2000 | | $(-\infty, 4.56)$ | $(-\infty, 4.44)$ | $(-\infty, 4.59)$ | $(-\infty, 4.47)$ | $(-\infty, 4.56)$ | $(-\infty, 4.44)$ |
| 4000 | | $(-\infty, 4.33)$ | $(-\infty, 4.27)$ | $(-\infty, 4.44)$ | $(-\infty, 4.28)$ | $(-\infty, 4.33)$ | $(-\infty, 4.27)$ |
| | | | | | | | |
| 500 | Upper 1-Sided | $(3.83, \infty)$ | $(3.32, \infty)$ | $(4.12, \infty)$ | $(3.55, \infty)$ | $(3.45, \infty)$ | $(3.03, \infty)$ |
| 1000 | | $(3.73, \infty)$ | $(3.48, \infty)$ | $(3.83, \infty)$ | $(3.56, \infty)$ | $(3.73, \infty)$ | $(3.48, \infty)$ |
| 2000 | | $(3.78, \infty)$ | $(3.67, \infty)$ | $(3.81, \infty)$ | $(3.69, \infty)$ | $(3.78, \infty)$ | $(3.67, \infty)$ |
| 4000 | | $(3.80, \infty)$ | $(3.74, \infty)$ | $(3.81, \infty)$ | $(3.75, \infty)$ | $(3.80, \infty)$ | $(3.74, \infty)$ |
| | | | | | | | |
| 500 | Symmetrical | (3.55,6.09) | (3.31.6.33) | (3.79,6.71) | (3.54,6.96) | (3.20,5.41) | (3.02,6.00) |
| 1000 | | (3.56,5.05) | (3.46,5.15) | (3.65,5.20) | (3.54,5.32) | (3.56,5.05) | (3.46,5.15) |
| 2000 | | (3.67,4.67) | (3.63,4.71) | (3.70,4.70) | (3.66,4.69) | (3.67,4.67) | (3.63,4.71) |
| 4000 | | (3.72,4.40) | (3.71,4.41) | (3.73,4.41) | (3.72,4.43) | (3.72,4.40) | (3.72,4.41) |

Table B.4.4: Logit Confidence Intervals for $\theta_{0n,2}$: Nominal level 0.90 $p_n = n/10$

| $n$ | Interval | Oracle Model Asymp. | Oracle Model Boot. | Pseudo Oracle Model Asymp., $C_n = 1$ | Pseudo Oracle Model Boot., $C_n = 1$ | Pseudo Oracle Model Asymp. $C_n = \log\log p_n$ | Pseudo Oracle Model Boot., $C_n = \log\log p_n$ |
|---|---|---|---|---|---|---|---|
| 500 | Lower 1-Sided | $(-\infty, -1.27)$ | $(-\infty, -1.04)$ | $(-\infty, -1.33)$ | $(-\infty, -1.10)$ | $(-\infty, -1.43)$ | $(-\infty, -1.20)$ |
| 1000 | | $(-\infty, -1.30)$ | $(-\infty, -1.20)$ | $(-\infty, -1.31)$ | $(-\infty, -1.20)$ | $(-\infty, -1.30)$ | $(-\infty, -1.20)$ |
| 2000 | | $(-\infty, -1.35)$ | $(-\infty, -1.30)$ | $(-\infty, -1.35)$ | $(-\infty, -1.30)$ | $(-\infty, -1.35)$ | $(-\infty, -1.30)$ |
| 4000 | | $(-\infty, -1.38)$ | $(-\infty, -1.36)$ | $(-\infty, -1.38)$ | $(-\infty, -1.36)$ | $(-\infty, -1.38)$ | $(-\infty, -1.36)$ |
| | | | | | | | |
| 500 | Upper 1-Sided | $(-2.29, \infty)$ | $(-2.03, \infty)$ | $(-2.37, \infty)$ | $(-2.10, \infty)$ | $(-2.48, \infty)$ | $(-2.21, \infty)$ |
| 1000 | | $(-1.93, \infty)$ | $(-1.83, \infty)$ | $(-1.94, \infty)$ | $(-1.84, \infty)$ | $(-1.93, \infty)$ | $(-1.83, \infty)$ |
| 2000 | $(-1.67, \infty)$ | $(-1.77, \infty)$ | $(-1.73, \infty)$ | $(-1.77, \infty)$ | $(-1.73, \infty)$ | $(-1.77, \infty)$ | $(-1.73, \infty)$ |
| 4000 | | $(-1.67, \infty)$ | $(-1.65, \infty)$ | $(-1.67, \infty)$ | $(-1.65, \infty)$ | $(-1.67, \infty)$ | $(-1.65, \infty)$ |
| | | | | | | | |
| 500 | Symmetrical | (-2.44,-1.12) | (-2.53,-1.03) | (-2.51,-1.18) | (-2.61,-1.08) | (-2.62,-1.28) | (-2.72,-1.19) |
| 1000 | | (-2.02,-1.21) | (-2.06,-1.18) | (-2.03,-1.22) | (-2.07,-1.18) | (-2.02,-1.21) | (-2.05,-1.76) |
| 2000 | | (-1.83,-1.29) | (-1.84,-1.27) | (-1.83,-1.29) | (-1.84,-1.28) | (-1.83,-1.29) | (-1.84,-1.27) |
| 4000 | | (-1.72,-1.34) | (-1.72,-1.33) | (-1.72,-1.34) | (-1.72,-1.33) | (-1.72,-1.34) | (-1.72,-1.33) |

Table B.4.5: Logit Confidence Intervals for $\theta_{0n,2}$: Nominal level 0.90 $p_n = n/2$

| $n$ | Interval | Oracle Model Asymp. | Oracle Model Boot. | Pseudo Oracle Model Asymp., $C_n = 1$ | Pseudo Oracle Model Boot., $C_n = 1$ | Pseudo Oracle Model Asymp. $C_n = \log\log p_n$ | Pseudo Oracle Model Boot., $C_n = \log\log p_n$ |
|---|---|---|---|---|---|---|---|
| 500 | Lower 1-Sided | $(-1.26, \infty)$ | $(-\infty, -1.03)$ | $(-\infty, -1.44)$ | $(-\infty, -1.18)$ | $(-\infty, -1.62)$ | $(-\infty, -1.39)$ |
| 1000 | | $(-1.29, \infty)$ | $(-\infty, -1.19)$ | $(-\infty, -1.31)$ | $(-\infty, -1.20)$ | $(-\infty, -1.30)$ | $(-\infty, -1.20)$ |
| 2000 | | $(-\infty, -1.33)$ | $(-\infty, -1.29)$ | $(-\infty, -1.34)$ | $(-\infty, -1.29)$ | $(-\infty, -1.33)$ | $(-\infty, -1.29)$ |
| 4000 | | $(-\infty, -1.38)$ | $(-\infty, -1.36)$ | $(-\infty, -1.38)$ | $(-\infty, -1.36)$ | $(-\infty, -1.38)$ | $(-\infty, -1.36)$ |
| | | | | | | | |
| 500 | Upper 1-Sided | $(-2.28, \infty)$ | $(-2.02, \infty)$ | $(-2.55, \infty)$ | $(-2.24, \infty)$ | $(-2.67, \infty)$ | $(-2.40, \infty)$ |
| 1000 | | $(-1.92, \infty)$ | $(-1.82, \infty)$ | $(-1.95, \infty)$ | $(-1.84, \infty)$ | $(-1.93, \infty)$ | $(-1.83, \infty)$ |
| 2000 | | $(-1.76, \infty)$ | $(-1.71, \infty)$ | $(-1.77, \infty)$ | $(-1.72, \infty)$ | $(-1.76, \infty)$ | $(-1.71, \infty)$ |
| 4000 | | $(-1.67, \infty)$ | $(-1.65, \infty)$ | $(-1.68, \infty)$ | $(-1.66, \infty)$ | $(-1.67, \infty)$ | $(-1.65, \infty)$ |
| | | | | | | | |
| 500 | Symmetrical | (-2.43,-1.11) | (-2.53,-1.01) | (-2.72,-1.28) | (-2.83,-1.17) | (-2.82,-1.47) | (-2.92,-1.37) |
| 1000 | | (-2.01,-1.20) | (-2.05,-1.17) | (-2.04,-1.22) | (-2.07,-1.18) | (-2.02,-1.21) | (-2.06,-1.18) |
| 2000 | | (-1.82,-1.27) | (-1.83,-1.26) | (-1.83,-1.28) | (-1.83,-1.27) | (-1.82,-1.27) | (-1.83,-1.26) |
| 4000 | | (-1.72,-1.34) | (-1.72,-1.33) | (-1.72,-1.34) | (-1.72,-1.34) | (-1.72,-1.33) | (-1.72,-1.33) |

Table B.4.6: Logit Confidence Intervals for $\theta_{0n,2}$: Nominal level 0.90 $p_n = 3n/4$

| $n$ | Interval | Oracle Model Asymp. | Oracle Model Boot. | Pseudo Oracle Model Asymp., $C_n = 1$ | Pseudo Oracle Model Boot., $C_n = 1$ | Pseudo Oracle Model Asymp. $C_n = \log\log p_n$ | Pseudo Oracle Model Boot., $C_n = \log\log p_n$ |
|---|---|---|---|---|---|---|---|
| 500 | Lower 1-Sided | $(-\infty,-1.3)$ | $(-\infty,-1.07)$ | $(-\infty,-1.52)$ | $(-\infty,-1.25)$ | $(-\infty,-1.58)$ | $(-\infty,-1.36)$ |
| 1000 | | $(-\infty,-1.29)$ | $(-\infty,-1.19)$ | $(-\infty,-1.33)$ | $(-\infty,-1.22)$ | $(-\infty,-1.29)$ | $(-\infty,-1.19)$ |
| 2000 | | $(-\infty,-1.35)$ | $(-\infty,-1.31)$ | $(-\infty,-1.36)$ | $(-\infty,-1.32)$ | $(-\infty,-1.35)$ | $(-\infty,-1.31)$ |
| 4000 | | $(-\infty,-1.37)$ | $(-\infty,-1.35)$ | $(-\infty,-1.38)$ | $(-\infty,-1.36)$ | $(-\infty,-1.37)$ | $(-\infty,-1.35)$ |
| | | | | | | | |
| 500 | Upper 1-Sided | $(-2.35,\infty)$ | $(-2.07,\infty)$ | $(-2.71,\infty)$ | $(-2.34,\infty)$ | $(-2.62,\infty)$ | $(-2.35,\infty)$ |
| 1000 | | $(-1.92,\infty)$ | $(-1.82,\infty)$ | $(-1.98,\infty)$ | $(-1.87,\infty)$ | $(-1.92,\infty)$ | $(-1.82,\infty)$ |
| 2000 | | $(-1.78,\infty)$ | $(-1.74,\infty)$ | $(-1.80,\infty)$ | $(-1.75,\infty)$ | $(-1.78,\infty)$ | $(-1.74,\infty)$ |
| 4000 | | $(-1.67,\infty)$ | $(-1.65,\infty)$ | $(-1.67,\infty)$ | $(-1.65,\infty)$ | $(-1.67,\infty)$ | $(-1.65,\infty)$ |
| | | | | | | | |
| 500 | Symmetrical | (-2.50,-1.16) | (-2.60,-1.06) | (-2.88,-1.35) | (-2.99,-1.24) | (-2.77,-1.43) | (-2.86,-1.34) |
| 1000 | | (-2.01,-1.20) | (-2.05,-1.17) | (-2.07,-1.23) | (-2.11,-1.20) | (-2.01,-1.20) | (-2.05,-1.17) |
| 2000 | | (-1.84,-1.29) | (-1.86,-1.28) | (-1.86,-1.30) | (-1.87,-1.29) | (-1.84,-1.29) | (-1.86,-1.28) |
| 4000 | | (-1.71,-1.33) | (-1.71,-1.33) | (-1.71,-1.34) | (-1.72,-1.33) | (-1.71,-1.33) | (-1.71,-1.33) |